\documentclass[aps,pra,twocolumn,10pt,superscriptaddress]{revtex4-2}
\pdfoutput=1

\usepackage[final]{graphicx}
\usepackage{times,bbm,amsmath,amssymb,amsthm}
\usepackage{epsfig,color}
\usepackage{xcolor}
\usepackage[colorlinks=true,citecolor=blue,linkcolor=blue]{hyperref}
\hypersetup{
   allcolors  = {blue},
}
\usepackage{cleveref}
\usepackage{subfiles}
\usepackage{float}
\usepackage{verbatim}
\usepackage[greek,english]{babel}
\usepackage{thumbpdf,enumerate}
\usepackage{booktabs}
\usepackage{sidecap}
\usepackage[scaled=.8]{couriers}    
\usepackage{pstricks}
\usepackage{multirow}
\usepackage{placeins}
\usepackage{relsize}  
\usepackage{pst-grad,bm}
\usepackage{epigraph}
\usepackage{booktabs}
\usepackage{bm}

\usepackage{soul}
\usepackage{ulem} 
\usepackage{acronym}
\usepackage{physics}
\usepackage{tikz}
\newsavebox{\imagebox}

\begin{document}

\title{Experimental verification of Threshold Quantum State Tomography on a fully-reconfigurable photonic integrated circuit}

\author{Eugenio Caruccio}
\affiliation{Dipartimento di Fisica, Sapienza Universit\`{a} di Roma, Piazzale Aldo Moro 5, I-00185 Roma, Italy}

\author{Diego Maragnano}
\affiliation{Dipartimento di Fisica, Universit\`{a} di Pavia, via Bassi 6, 27100 Pavia, Italy}

\author{Giovanni Rodari}
\affiliation{Dipartimento di Fisica, Sapienza Universit\`{a} di Roma, Piazzale Aldo Moro 5, I-00185 Roma, Italy}

\author{Davide Picus}
\affiliation{Dipartimento di Fisica, Sapienza Universit\`{a} di Roma, Piazzale Aldo Moro 5, I-00185 Roma, Italy}

\author{Giovanni Garberoglio}
\affiliation{European Centre for Theoretical Studies in Nuclear Physics and Related Areas (ECT*, Fondazione Bruno Kessler); Villa Tambosi, Strada delle Tabarelle 286, I-38123 Villazzano (TN), Italy}

\author{Daniele Binosi}
\affiliation{European Centre for Theoretical Studies in Nuclear Physics and Related Areas (ECT*, Fondazione Bruno Kessler); Villa Tambosi, Strada delle Tabarelle 286, I-38123 Villazzano (TN), Italy}

\author{Riccardo Albiero}
\affiliation{Istituto di Fotonica e Nanotecnologie, Consiglio Nazionale delle Ricerche (IFN-CNR), 
Piazza Leonardo da Vinci, 32, I-20133 Milano, Italy}

\author{Niki Di Giano}
\affiliation{Dipartimento di Fisica, Politecnico di Milano, Piazza Leonardo da Vinci 32, 20133 Milano, Italy}
\affiliation{Istituto di Fotonica e Nanotecnologie, Consiglio Nazionale delle Ricerche (IFN-CNR), 
Piazza Leonardo da Vinci, 32, I-20133 Milano, Italy}

\author{Francesco Ceccarelli}
\affiliation{Istituto di Fotonica e Nanotecnologie, Consiglio Nazionale delle Ricerche (IFN-CNR), 
Piazza Leonardo da Vinci, 32, I-20133 Milano, Italy}

\author{Giacomo Corrielli}
\affiliation{Istituto di Fotonica e Nanotecnologie, Consiglio Nazionale delle Ricerche (IFN-CNR), 
Piazza Leonardo da Vinci, 32, I-20133 Milano, Italy}

\author{Nicol\`o Spagnolo}
\affiliation{Dipartimento di Fisica, Sapienza Universit\`{a} di Roma, Piazzale Aldo Moro 5, I-00185 Roma, Italy}

\author{Roberto Osellame}
\email{roberto.osellame@cnr.it}
\affiliation{Istituto di Fotonica e Nanotecnologie, Consiglio Nazionale delle Ricerche (IFN-CNR), 
Piazza Leonardo da Vinci, 32, I-20133 Milano, Italy}

\author{Maurizio Dapor}
\email{dapor@ectstar.eu}
\affiliation{European Centre for Theoretical Studies in Nuclear Physics and Related Areas (ECT*, Fondazione Bruno Kessler); Villa Tambosi, Strada delle Tabarelle 286, I-38123 Villazzano (TN), Italy}

\author{Marco Liscidini}
\email{marco.liscidini@unipv.it}
\affiliation{Dipartimento di Fisica, Universit\`{a} di Pavia, via Bassi 6, 27100 Pavia, Italy}

\author{Fabio Sciarrino}
\email{fabio.sciarrino@uniroma1.it}
\affiliation{Dipartimento di Fisica, Sapienza Universit\`{a} di Roma, Piazzale Aldo Moro 5, I-00185 Roma, Italy}

%---------------------------------------------

\begin{abstract}
Reconstructing the state of a complex quantum system represents a pivotal task for all quantum information applications, both for characterization purposes and for verification of quantum protocols. Recent technological developments have shown the capability of building quantum systems with progressively larger number of qubits in different platforms. The standard approach based on quantum state tomography, while providing a method to completely characterize an unknown quantum state, requires a number of measurements that scales exponentially with the number of qubits. Other methods have been subsequently proposed and tested to reduce the number of measurements, or to focus on specific properties of the output state rather than on its complete reconstruction. Here, we show experimentally the application of an approach, called threshold quantum state tomography, in an advanced hybrid photonic platform with states up to $n=4$ qubits. This method does not require a priori knowledge on the state, and selects only the informative projectors starting from the measurement of the density matrix diagonal. We show the effectiveness of this approach in a photonic platform, showing that a consistent reduction in the number of measurement is obtained while reconstructing relevant states for quantum protocols, with only very limited loss of information. The advantage of this protocol opens perspective of its application in larger, more complex, systems.
\end{abstract}
%---------------------------------------------
\maketitle
%---------------------------------------------

\section{Introduction}

The ability to accurately characterize a quantum state is a fundamental requirement for the development of quantum technologies. In quantum computing, quantum communication, and quantum metrology, precise knowledge of the state of a system enables the verification of quantum protocols, the validation of experimental implementations, and the assessment of device performance. However, obtaining a complete description of a quantum state is a challenging task. A general quantum state is represented by its density matrix, which encodes all the relevant statistical properties of the system. In principle, knowing this matrix provides access to any observable quantity. Yet, reconstructing the density matrix from experimental data using Quantum State Tomography (QST) requires performing a series of measurements, whose number scales exponentially with the system size \cite{james2001measurement, thew2002qudit, d2003quantum,altepeter2005photonic}. 

As quantum platforms advance towards higher-dimensional states and larger numbers of qubits, dealing with this exponential scaling is crucial in quantum information science. While the constraints of laboratory resources, including measurement time and computational capabilities, limit the feasibility of QST, alternative strategies have been developed to mitigate this issue. Compressed sensing QST, for example, reduces the number of required measurements by leveraging prior assumptions on the rank of the density matrix \cite{gross2010quantum, flammia2012quantum, tonolini2014reconstructing}. Bayesian QST, in contrast, introduces probabilistic models to optimize measurement selection \cite{blume2010optimal, lu2018quantum, lu2022bayesian}. Other approaches, such as shadow tomography, aim to extract relevant information while avoiding full-state reconstruction \cite{aaronson2018shadow, huang2020predicting, struchalin2021experimental}. 

An alternative strategy is represented by threshold Quantum State Tomography (tQST) \cite{binosi2024tailor}. The core idea behind tQST is to exploit the structure of the density matrix to prioritize the most significant measurements. Specifically, it leverages the fact that the off-diagonal elements of a density matrix are constrained by the diagonal ones, allowing for a systematic reduction in the number of required measurements. The protocol follows a simple sequence: first, the diagonal elements of the density matrix are measured on the given computational basis; second, a threshold parameter is introduced to determine which off-diagonal elements are likely to contribute significantly; finally, only those selected elements are measured, and the density matrix is reconstructed based on this reduced dataset. By adjusting the threshold, tQST provides a tunable balance between measurement effort and reconstruction accuracy. This can lead to a significant reduction of measurements, particularly for systems where the density matrix is naturally sparse.

While tQST has been successfully demonstrated in superconducting qubit systems \cite{binosi2024tailor}, extending its application to photonic platforms represents an important step toward broader applicability, as these systems offer unique advantages in terms of scalability, coherence properties, and compatibility with integrated architectures \cite{wang2020integrated, giordani2023integrated}. In this work, we experimentally validate tQST in a photonic system based on single photons generated from a quantum dot source and manipulated within a femtosecond-laser-written integrated circuit. This platform combines high-performance photon generation with fully reconfigurable photonic processing capabilities, making it well-suited for scalable quantum information processing. Photonic qubits encoded in spatial and polarization modes provide a flexible framework for implementing high-dimensional quantum states, enabling an exploration of tQST in a regime distinct from previous demonstrations in superconducting circuits.

The rest of this manuscript is structured as follows. In Section II, we provide the theoretical background of QST and introduce the principles of tQST, outlining its advantages and operational framework. In Section III, we describe the experimental setup, detailing the photonic platform used to implement the protocol. Section IV presents the results of our experimental validation, analyzing its effectiveness for different classes of quantum states. Finally, in Section V, we summarize our findings and discuss potential directions for future research.

%---------------------------------------------

\section{Theoretical Background}

QST aims to reconstruct the representation of a quantum state by measuring a sufficiently large number of observables. In this work, we consider the density matrix as the representation of quantum states, \textit{i.e.}, a trace-one, Hermitian, and positive semi-definite matrix. The number of required observables is determined by the dimension of the density matrix. Specifically, for a system of \( n \) qubits, in principle \( 4^n -1 \) observables are necessary, corresponding to the number of independent real parameters characterizing the density matrix. One can perform the measurements in any order, and subsequently process the acquired data using appropriate statistical inference methods, such as maximum likelihood estimation or Bayesian mean estimation \cite{james2001measurement, blume2010optimal}.

The properties of density matrices, particularly positive semi-definiteness, impose constraints on their off-diagonal elements \( \rho_{ij} \), specifically that \( |\rho_{ij}| \leq \sqrt{\rho_{ii} \rho_{jj}} \). Consequently, measuring the diagonal elements provides information about the off-diagonal ones. For example, if \( \rho_{ii} \) is zero, then all elements in the \( i \)-th row and column of \( \rho \) must also be zero. Similarly, if \( \rho_{ii} \) and \( \rho_{jj} \) are nonzero but small relative to other diagonal elements, the modulus of \( \rho_{ij} \) will also be small.

\begin{figure}[ht!]
    \centering
    \includegraphics[width=0.49\textwidth]{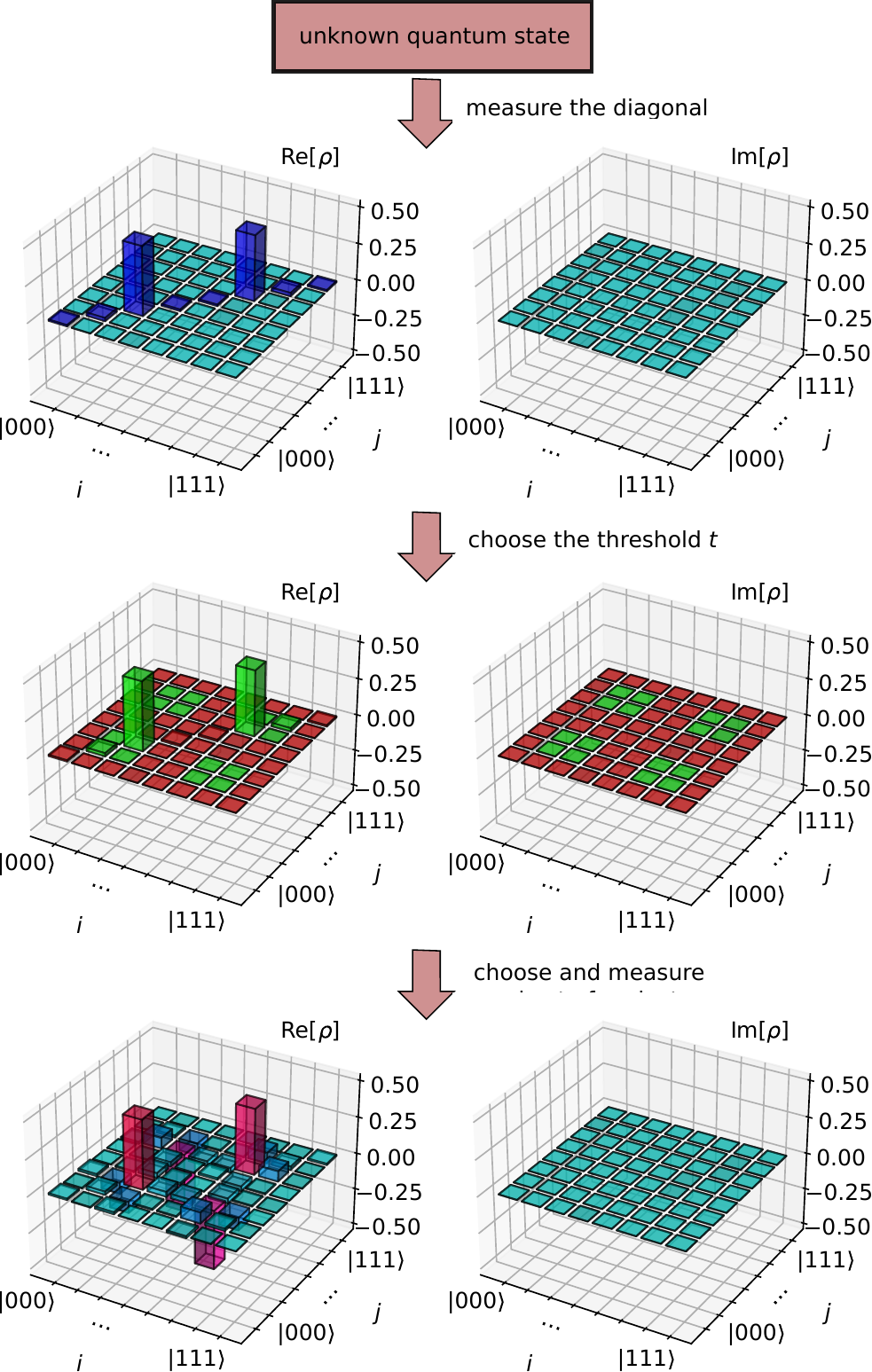}
    \caption{\textbf{Conceptual visualization of the tQST protocol.} Starting from an unknown quantum state, the protocol requires reconstructing the diagonal elements $\rho_{ii}$ of the density matrix, which is done by measuring the state in the computational basis (upper panel, blue entries). Then, after a threshold $t$ is appropriately set, the protocol chooses the projectors to be measured to gain information only on those off-diagonal elements $\rho_{ij}$ corresponding to the condition $\sqrt{\rho_{ii} \rho_{jj}} \geq t$ (middle panel, green entries). From those measurements the density matrix can be then reconstructed via maximum likelihood (lower panel)}
    \label{fig:conceptual}
\end{figure}

These observations form the basis of the tQST protocol, which we now describe \cite{binosi2024tailor}. The protocol (schematically shown in Fig. \ref{fig:conceptual}) consists of the following steps: (i) we first directly measure the diagonal elements \( \{\rho_{ii}\} \) of the density matrix by projecting onto the elements of the chosen computational basis; (ii) we choose a threshold \( t \), and using the information from \( \{\rho_{ii}\} \), the off-diagonal elements \( \rho_{ij} \) satisfying \( \sqrt{\rho_{ii} \rho_{jj}} \geq t \) are identified; (iii) a set of projectors providing information on these selected \( \rho_{ij} \) elements is constructed, and we perform only these measurements; (iv) finally, we process the measurement results using statistical inference techniques and reconstruct the density matrix.

Several key aspects of the tQST protocol warrant discussion. The resources required to complete the experiment are predetermined once the threshold is set in the tQST protocol. In contrast, adaptive approaches select each measurement based on the outcome of the previous one \cite{ahn2019adaptive, ahn2019adaptive_numerical}. tQST does not make any prior assumptions about the state, unlike some other methods that reduce the number of measurements or improve scaling by assuming a specific structure of the quantum state \cite{gross2010quantum, baumgratz2013scalable}. The threshold \( t \) serves to control the resources required for the protocol, such as the time needed for the measurements and the computational power for data processing. By choosing \( t > 0 \), fewer resources may be needed compared to QST. The threshold \( t \) can be set by the user based on available resources. However, tQST suggests that the reduction in measurements is particularly significant for sparse density matrices. To relate \( t \) to matrix sparsity, we derive a formula for \( t \) based on the initial measurements, {\em i.e.}, the diagonal elements of the density matrix. 

To this end, we employ the Gini index \cite{hurley2009comparing}. Given \( \underline{c} = \left[ c_{(1)}, c_{(2)}, \dots, c_{(N)} \right] \) where \( c_{(1)} \le c_{(2)} \le \dots \le c_{(N)} \) and \( c_{(i)} \geq 0  ~ \forall  i \), the Gini index is defined as:
\begin{equation}\label{eqn: gini}
    \text{GI}(\boldsymbol{c}) = 1 -2 \sum_{k=1}^N\frac{c_k}{\| \boldsymbol{c} \|_1}\left( \frac{N-k+\frac{1}{2}}{N} \right),
\end{equation}
where $\| \boldsymbol{c}\|_1 = \sum_{i=1}^N c_i$ identifies the usual Euclidean $1$-norm. It holds \( 0 \leq \text{GI} \left( \boldsymbol{c} \right) \leq 1 - 1/N \), where the lower bound is attained where all the $c_{(i)}$ are equal, and the higher bound denotes maximum inequality. We adapt this definition to make the Gini index a suitable threshold for tQST. Based on the protocol, the relevant vector for computing the Gini index is the diagonal of the density matrix, where \( N = 2^n \). The threshold is then set as:
\begin{equation}\label{eq:threshold_gini}
     t = \|{\boldsymbol \rho}\|_1 ~  \frac{\text{GI} \left( \boldsymbol{\rho} \right)}{2^n-1},
\end{equation}
with \( \boldsymbol{\rho} = \left( \rho_{11}, \rho_{22}, \dots, \rho_{NN} \right) \). 

\section{Experimental apparatus}

\begin{figure*}[ht!]
    \centering
    \includegraphics[width=0.99\linewidth]{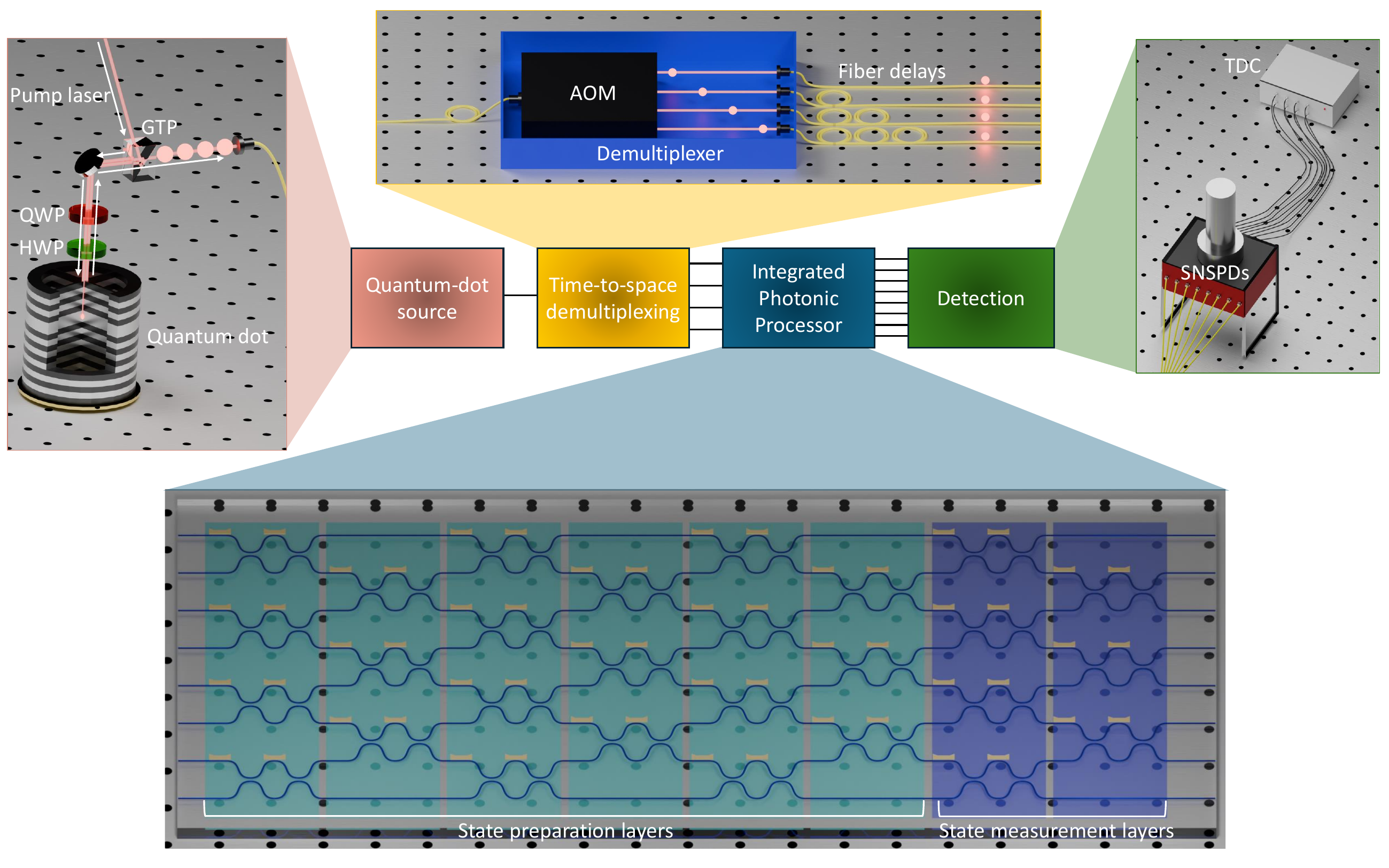}
    \caption{\textbf{Architecture of the hybrid-photonic platform employed in the experiment.} The experimental apparatus, based on a hybrid approach, has been used to generated multi-qubit states and reconstruct them via the tQST technique. A pulsed laser, manipulated with a pulse-shaping stage, is used to excite a QD single-photon source employing a resonance fluorescence (RF) optical excitation technique, comprising a Glan--Thompson polarizer (GTP), a half wave plate (HWP) and a quarted wave plate (QWP). Then, the single-photon stream emitted by the QD source is interfaced with the demultiplexing module, based on an acousto-optic modulator (AOM), in order to obtain the multi-photon state required for the implementation of multi-photon protocols. The output resource is then interfaced with an 8-mode fully-reconfigurable integrated photonic processor, fabricated via the femtosecond-laser-writing technology, with internal structure allowing for universal operation. The first six layers (cyan) of RBSs are used to perform the state generation process in dual rail encoding, while the remaining two layers (blue) are used to measure the qubits in different bases. At the output of the setup, photon detection is carried out via superconducting-nanowire single-photon detectors (SNSPDs). Photon detection events, corresponding to $n$-fold coincidences, are registered via a time-to-digital converter (TDC).}
    \label{fig:gensetup}
\end{figure*}

The experimental setup comprises a state-of-the-art hybrid photonic architecture, interconnecting a diverse array of technologies \cite{elshaari2020hybrid} and tailored for photon-based multi-photon experiments \cite{rodari2024semi,rodari2024experimental}. More specifically, it comprises three subsequent stages, as depicted in Fig. \ref{fig:gensetup}. Namely, the photon resources are generated via a single-photon source based on quantum-dot (QD) technology \cite{michler2017quantum,senellart2017high,heindel2023quantum,esmann2024solid}, excited at a repetition rate of $158$ MHz in the resonance fluorescence (RF) regime \cite{somaschi2016near,ollivier2020reproducibility,nowak2014deterministic,gazzano2013bright,senellart2017high}. Then, a time-to-spatial demultiplexing module (DMX) based on an acousto-optical modulator \cite{rodari2024semi,rodari2024experimental, hansen2023single,pont2022quantifying,pont2022high} converts the initial train of single photons, emitted at a fixed time interval, into sets of photons distributed in several spatial modes. This procedure permits to generate a multi-photon resource, which can be injected into the input ports of the experiment. The pairwise indistinguishability between the photons comprising the generated four-photon state is guaranteed by finely tuned free-space delay lines and polarization paddle controllers. The multi-photon states generated with such an approach are then injected into an $8$-mode fully reconfigurable integrated photonic processor. Specifically, the present device was fabricated via the femtosecond laser-writing technique \cite{corrielli2021femtosecond}, and its internal structure is based on an interferometric mesh of 28 Mach--Zehnder unit cells, acting as Reconfigurable Beam Splitter (RBS), arranged according to the universal rectangular geometry described in \cite{clements2016optimal}. Moreover, the integrated processor employed here features polarization-independent operation \cite{pentangelo2024high}. Its reconfigurability, induced by thermal-phase shifters \cite{flamini2015thermally,ceccarelli2019thermal,ceccarelli2020low, albiero2022, ceccarelli2024integrated}, is suitably controlled via current controllers which can be set to implement a given chosen unitary transformation $U$ \cite{pentangelo2024high}. Finally, at the output of the setup, the photons are collected from the integrated processor and then sent to a suitable detection system comprising high-efficiency superconducting nanowire single-photon detectors (SNSPDs), paired with a time-to-digital converter (TDC) used to discriminate and register $n$-fold coincidence events.

\section{Experimental Results}

Hereafter, we will present the results of the photonic experiment aiming to verify and validate the tQST technique \cite{binosi2024tailor}, by analyzing different classes of states, corresponding to different level of sparsities of the density matrices. The results are benchmarked by considering the reconstruction via the tQST approach, with respect to QST which in our case is implemented via tQST with $t=0$. The latter method uses $4^{n}$ projectors, chosen to be a tomographically-complete set \cite{james2001measurement}. We will thus refer hereafter to tQST with threshold $t=0$ as QST.

\subsection{Experimental implementation}
Quantum states are generated in the photon path degree of freedom by harnessing linear optics elements, such as beam splitters and phase shifters, and considering $n$-fold coincidence events measured in post-selection on a subset of all possible combinations of $n$ photons in $m$ modes.  Namely, qubit states are encoded via the \textit{dual rail logic} \cite{kok2007linear} in the photon spatial modes. Thus, the generation of a $n$-qubit quantum state is obtained non-deterministically by considering an optical interferometer with an even number $m=2n$ of modes as split into $n$ dual rails, that is, adjacent pairs of spatial modes labeled with states $\ket{0}$ and $\ket{1}$ from top to bottom (see Fig. \ref{fig:gensetup}). Overall, when $n$ photons are injected in the circuit, only the output combinations of photons distributed over the $m$ modes, such that each dual rail is occupied by one and only one photon, are considered valid for state generation, while the others are discarded via a post-selection process.

Controlled generation of the $n$-qubit state and its subsequent reconstruction via tomographic measurements both take place in the aforementioned reconfigurable integrated photonic processor. As shown in Fig. \ref{fig:gensetup} and in the Supplementary Material, the first six layers of the interferometer are suitably configured to generate a chosen state. In such a view, the generation of a $n$-qubit state is obtained by mapping an initial $n$-photon state, via a suitable linear optical transformation, into a set of $2n$ modes and by applying a suitable post-selection procedure upon the detection of an $n$-photon coincidence event in mode combinations satisfying the dual rail logic. Conversely, the remaining two layers of the circuit are dedicated to the implementation of measurement projectors. More specifically, this relies on the possibility of implementing any arbitrary single-qubit transformation via a single elementary cell. Thus, by appropriately programming the transformation of a RBS acting on the two modes of each qubit, it is possible to implement any projective measurement on the output state. By applying the appropriate sequence of measurements according to the chosen reconstruction algorithm (either QST or tQST in this experiment), one can measure the observables needed to retrieve the initial quantum state.

\begin{figure*}[ht!]
    \centering
    \includegraphics[width=0.99\textwidth]{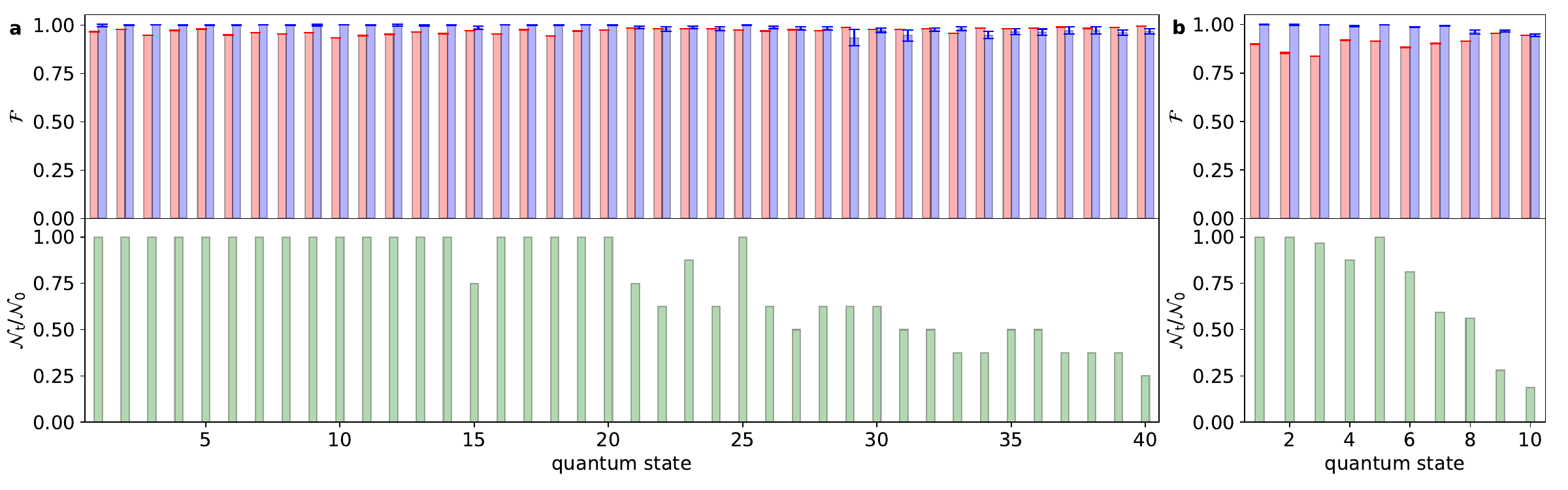}
    \caption{\textbf{Benchmarking the tQST procedure through $2$- and $3$-qubit quantum state reconstruction.} The tQST approach is tested by generating and analyzing a set of quantum states with equally spaced Gini index, ranging between its minimum and maximum possible value. \textbf{a}, Scenario with $n=2$ qubits, tested with 40 different states. \textbf{b}, Scenario with $n=3$ qubits, tested with 10 different states. In the upper plots, we report the fidelities $\mathcal{F}^{(n)}_{\mathrm{0,m}}$ (with $n=2,3$) between the state reconstructed via QST and the theoretical model taking into account experimental imperfections (red bars), and the fidelities $\mathcal{F}^{(n)}_{\mathrm{0,t}}$  (with $n=2,3$) between the state reconstructed via QST and tQST choosing the threshold according to the Gini index (blue bars). In the lower plots, we report the ratio $\mathcal{N}_{\mathrm{t}}/\mathcal{N}_{0}$ (green bars) between the number of projectors $\mathcal{N}_{\mathrm{t}}$ selected to be measured by the tQST approach with threshold according to Eq. \eqref{eq:threshold_gini}, and the number of projectors $\mathcal{N}_{0}$ corresponding to the QST approach. The states are ordered on the $x$-axis according to the associated Gini indexes.}
    \label{fig:resultsgini}
\end{figure*}

The main figure of merit to quantify the efficacy of the tQST approach is given by the fidelity between the density matrix derived using tQST and the one obtained via the QST procedure ($\mathcal{F}_{\mathrm{0,t}}$). This parameter corresponds to the overlap between the two reconstructions, and thus quantifies the amount of information that is lost by reducing the number of projectors via the tQST approach. Values of $\mathcal{F}_{\mathrm{0,t}}$ close to one highlight that the tQST method provides an effective reconstruction of the state. A second set of relevant parameters are the purities $\mathcal{P}_{0}$ and $\mathcal{P}_{\mathrm{t}}$ of the reconstructed states. The purity $\mathcal{P}_{0}$ represents the intrinsic value for the reconstructed state, which is due to the different noise sources in the apparatus. Additionally, evaluation of $\mathcal{P}_{\mathrm{t}}$ and its comparison with $\mathcal{P}_{0}$ permits to verify how the purity of the reconstructed state is affected by the reduced number of projectors. Finally, the remaining relevant parameter is the fidelity between the density matrix derived using the QST procedure and the expected state from the apparatus ($\mathcal{F}_{\mathrm{0,m}}$). This expected state is obtained via numerical simulations that consider the main noise effects in the apparatus, due to non-ideal single-photon emission from the QD source \cite{pont2022high,pont2022quantifying,rodari2024experimental,rodari2024semi}, to losses and to an imperfect dialling of the unitary matrix on the integrated photonic processor. The noise effects that we consider are related to i) a non-zero multi-photon emission probability from the QD source, ii) a degree of partial distinguishability between the photons and iii) directional couplers of the elementary cell with a splitting ratio different from $50$:$50$ (more details can be found in the Supplementary Material). The parameters related to the different effects have been estimated by independent measurements. More specifically, noise source i) is characterized by a standard measurement of the second-order autocorrelation via a Hanbury~Brown--Twiss interferometer, which leads to a value of $g^{(2)} (0) \sim 0.01$. Conversely, partial distinguishability ii) can be estimated through measurement of the Hong--Ou--Mandel visibilities $V^{\mathrm{HOM}}$ between the photons emitted at different times which, averaged among all possible photon pairs, provided a value of $V^{\mathrm{HOM}} \sim 0.90$. Finally, the splitting ratios of the directional couplers iii) have been retrieved during the circuit calibration procedure.

\subsection{tQST validation for two and three-qubit states}

As a first step, we have implemented experimentally the tQST technique on a set of states for $n=2$ and $n=3$ qubits. More specifically, we have generated a set of random states for each dimension, and selected a subset of 40 states for $n=2$, and 10 states for $n=3$, where each set is characterized by equally-spaced values of the Gini index, from the minimum to the maximum value, calculated on the diagonal elements of the density matrices. This approach allows to test the performances of the tQST method by considering states covering uniformly the full range of possible degrees of sparsity in the diagonal. The chosen states are generated, according to the procedure above, in the dual rail logic, by using $n$ input photons and a portion of the circuit composed of $m=2n$ modes. For each value of the number of qubits, a number of RBSs equal to $N_{\text{RBS}}=m(m-1)/2$, which constitute a $m$-mode universal multiport interferometer, are exploited in the state preparation stage (more details on the layout are found in the Supplementary Material). Population and coherence elements of the density matrices are processed via the measurement layers, comprising single-qubit projectors.

\begin{figure*}[ht!]
    \centering
    \includegraphics[width=0.99\textwidth]{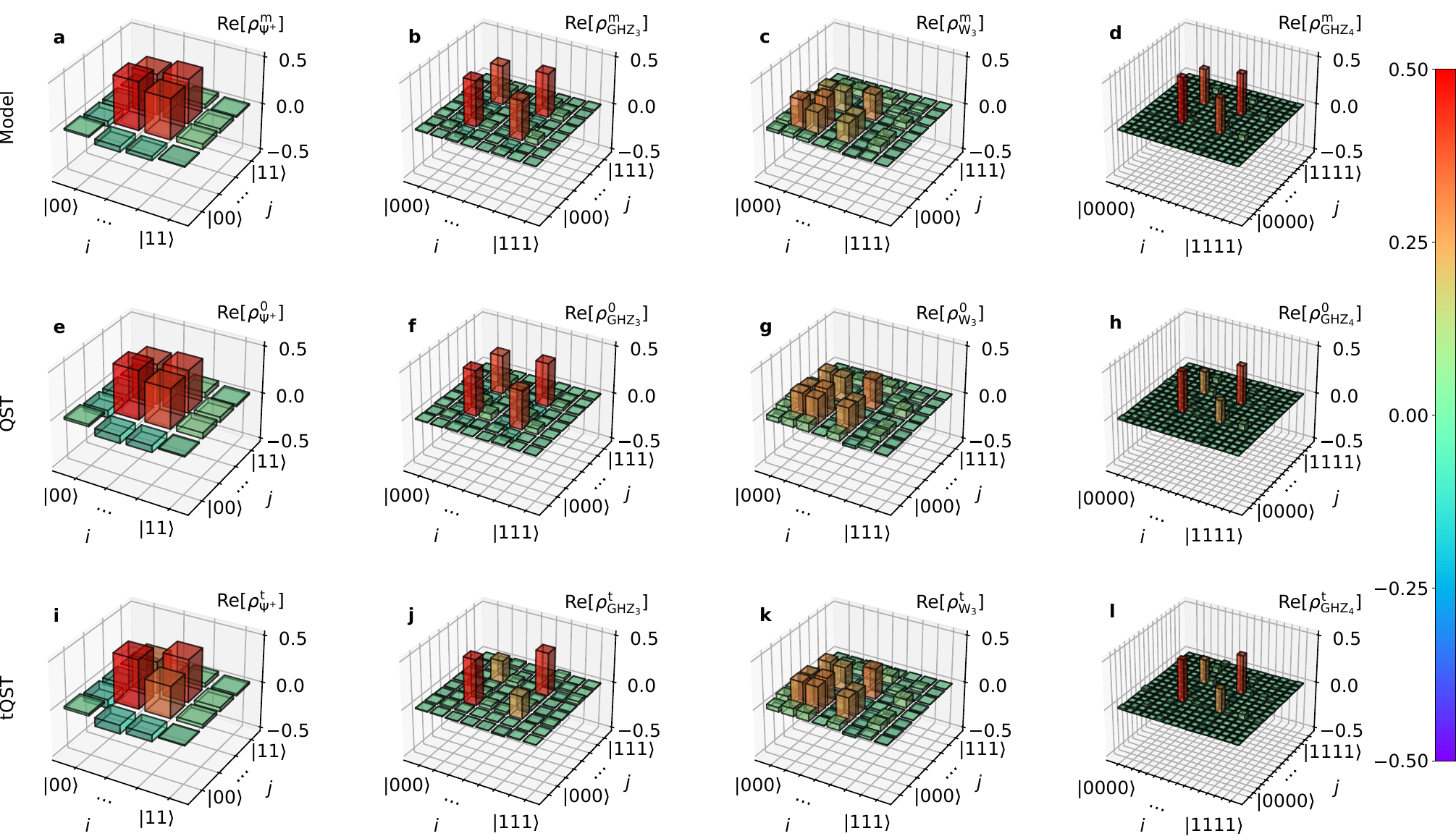}
    \caption{\textbf{Results of the state reconstruction process for maximally-entangled states.} Real part of the density matrices for the different tested maximally-entangled states, namely (from left to right) $\vert \Psi^{+} \rangle$ for $n=2$, $\vert \mathrm{GHZ}_{3} \rangle$ and $\vert \mathrm{W}_{3} \rangle$ for $n=3$ and $\vert \mathrm{GHZ}_{4} \rangle$ for $n=4$. In the first row (panels \textbf{a-d}) we report the expected density matrices ($\rho^{\mathrm{m}}_{\alpha}$) estimated from the model taking into account experimental imperfections. The second row (panels \textbf{e-h}) reports the corresponding experimentally reconstructed density matrices ($\rho^{0}_{\alpha}$) via the QST approach, while the third row (panels \textbf{i-l}) reports the density matrices ($\rho^{\mathrm{t}}_{\alpha}$) retrieved via the tQST approach with threshold chosen according to the Gini index. The index $\alpha$ labels the states as $\alpha = \Psi^{+}, \mathrm{GHZ}_{3}, \mathrm{W}_{3}, \mathrm{GHZ}_{4}$. On the right part of the figure, we report the colormap for the density matrix bars, equal for all plots \textbf{a}-\textbf{l}.
    }
    \label{fig:densitymatrices}
\end{figure*}

The results of the state reconstruction process with both QST and tQST are shown in Fig. \ref{fig:resultsgini}. For each analyzed state we report the fidelity $\mathcal{F}^{(n)}_{\mathrm{0,t}}$ between the density matrices obtained with the two approaches, and perform a joint analysis with the ratio $\mathcal{N}_{\mathrm{t}}/\mathcal{N}_{0}$ between the number of projectors used by tQST ($\mathcal{N}_{t}$) and the ones used by QST ($\mathcal{N}_{0}$). Here, the reconstruction with the tQST method is performed by choosing the threshold according to the Gini index [Eq. (\ref{eq:threshold_gini})]. We observe that, for states with the lowest sparsity values, the two techniques coincide given that the tQST approach requires measuring a tomographically-complete set of $4^{n}$ projectors. When measuring states with increasing sparsity in the diagonal elements, the tQST starts to be advantageous due to the measurement of a progressively lower number of projectors, as shown in the bottom panels of Fig. \ref{fig:resultsgini}. The obtained quality in the reconstructions for all states shows that tQST is effective in reducing the number of projectors while having only a very limited loss of information with respect to the QST approach, since all fidelities are found to be $\mathcal{F}^{(n)}_{\mathrm{0,t}} > 0.935$. Such advantage in reducing the number of projectors becomes more pronounced when increasing the state dimensionality, given the exponential increase in the number of measurements required by QST. As a final note, we observe that the reconstructed states show a good agreement with the theoretical expectations from the model (see also Supplementary Material for additional comparisons), as quantified by the average fidelities $\langle \mathcal{F}^{(2)}_{\mathrm{0,m}}\rangle = 0.969\pm 0.014$ for the $2$-qubit scenario, and $\langle \mathcal{F}^{(3)}_{\mathrm{0,m}} \rangle = 0.903\pm 0.035$, for the $3$-qubit case.

\subsection{tQST implementation for maximally-entangled states}

As a subsequent step to test the tQST approach, we have analyzed its application to reconstruct specific maximally-entangled states characterized by a density matrix comprising a large number of zero-valued elements. In this case, the tQST method is expected to maximize the advantage with respect to the QST approach. Indeed, in a noiseless scenario a large number of projectors correspond to zero-valued elements, and the protocol does not require their measurement to reconstruct the state. We have then tested different maximally-entangled states, such as $\ket{\Psi^{+}} = (\ket{01}+\ket{10})/\sqrt{2}$ for $n=2$ qubits, $\ket{\mathrm{GHZ}_3} = (\ket{010}+\ket{101})/\sqrt{2}$ and $\ket{\mathrm{W}_3} = (\ket{100}+\ket{010}+\ket{001})/\sqrt{3}$ for $n=3$ qubits, and $\ket{\mathrm{GHZ}_4} = (\ket{0101}+\ket{1010})/\sqrt{2}$ for $n=4$ qubits. The Bell state and the GHZ states are generated by following the post-selected approach of \cite{pont2022high}, while the W state is generated by exploiting the full reprogrammability of the device (more details on the configuration for the generation layout are found in the Supplementary Material). 

\begin{figure*}[ht!]
    \centering
    \includegraphics[width=0.99\textwidth]{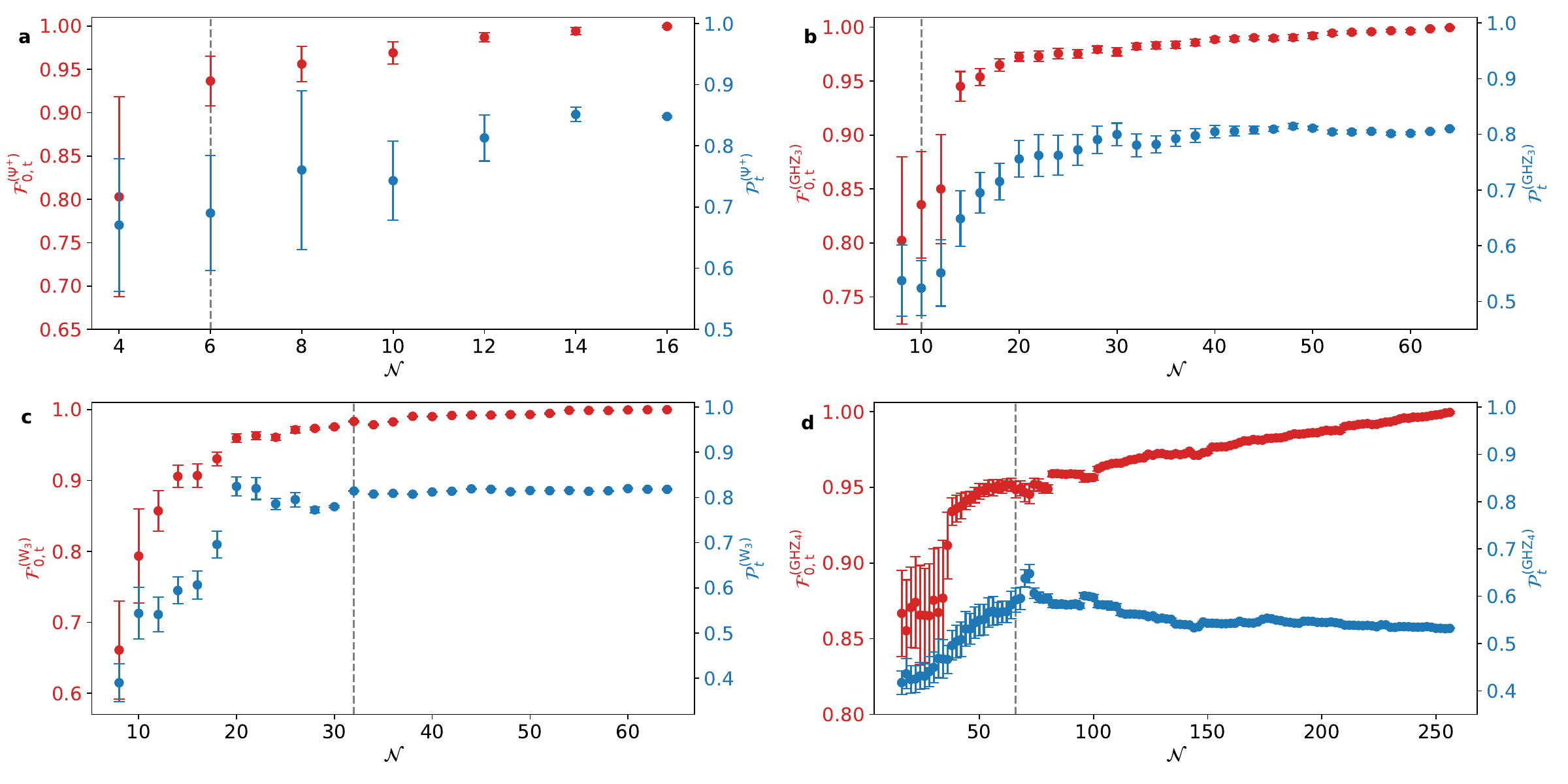}
    \caption{\textbf{Reconstruction of maximally-entangled states.} Red points: plots of the fidelities ($\mathcal{F}^{(\alpha)}_{\mathrm{0,t}}$) between the state reconstructed via QST, and the state reconstructed via tQST for different values of the number projectors $\mathcal{N}$, obtained by progressively lowering the value of the threshold $t$. Blue points: corresponding purities ($\mathcal{P}^{(\alpha)}_{\mathrm{t}}$) for the states reconstructed via tQST as a function of $\mathcal{N}$. The index $\alpha$ labels the states as $\alpha = \Psi^{+}, \mathrm{GHZ}_{3}, \mathrm{W}_{3}, \mathrm{GHZ}_{4}$. \textbf{a}, Plots for state $\vert \Psi^{+} \rangle$. \textbf{b}, Plots for state $\vert \mathrm{GHZ}_{3} \rangle$. \textbf{c}, Plots for state $\vert \mathrm{W}_{3} \rangle$. \textbf{d}, Plots for state $\vert \mathrm{GHZ}_{4} \rangle$. In all plots, the vertical dashed lines correspond to the number of projectors obtained using the threshold computed from the Gini index [Eq.~(\ref{eq:threshold_gini})].}
    \label{fig:fidelitiespurities}
\end{figure*}

The results for the reconstructed density matrices with the two methods, and their comparison with the model of the apparatus, are shown in Fig. \ref{fig:densitymatrices} (complementary analyses are reported in the Supplementary Material). We observe that, for all states, the reconstructed states with the tQST approach, using the Gini index to choose the threshold, are close to those obtained with QST. A more detailed analysis on the performances can be found in Tab. \ref{tab:maximally_fidelities} and in Fig. \ref{fig:fidelitiespurities}, where we have analyzed tQST by further using different number of projectors $\mathcal{N}$ with respect to the one indicated by the Gini threshold. 
\begin{table}[ht!]
\begin{tabular}{|c|c|c|c|}
\hline
State & $\mathcal{N}_{0}$ & $\mathcal{N}_{\mathrm{t}}$ & $\mathcal{F}^{(\alpha)}_{\mathrm{0,t}}$ \\
\hline
$\vert \Psi^{+} \rangle$ & 16 & 6 & $0.94 \pm 0.03$\\
$\vert \mathrm{GHZ}_{3} \rangle$ & 64 & 10 & $0.84 \pm 0.05$\\
$\vert \mathrm{W}_{3} \rangle$ & 64 & 32 & $0.9833 \pm 0.0007$\\
$\vert \mathrm{GHZ}_{4} \rangle$ & 256 & 66 & $0.948 \pm 0.005$ \\
\hline
\end{tabular}
\caption{\textbf{Reconstruction of maximally-entangled states.} Values of the number of projectors $\mathcal{N}_{\mathrm{t}}$ used by tQST with a value of the threshold chosen according to the Gini index, of the number of projectors $\mathcal{N}_{0}$ required by QST, and of the fidelities $\mathcal{F}^{(\alpha)}_{\mathrm{0,t}}$ between the two reconstruction techniques. The index $\alpha$ labels the states as $\alpha = \Psi^{+}, \mathrm{GHZ}_{3}, \mathrm{W}_{3}, \mathrm{GHZ}_{4}$.}
\label{tab:maximally_fidelities}
\end{table}
We start by analyzing the fidelity between QST and tQST $\mathcal{F}^{(\alpha)}_{\mathrm{0,t}}$ as a function of the number of projectors $\mathcal{N}$, with $\alpha = \Psi^{+}, \mathrm{GHZ}_{3}, \mathrm{W}_{3}, \mathrm{GHZ}_{4}$. We observe that, for all states, the initial reduction of the number of projectors in a regime above the value identified by the Gini index is accompanied by a small loss of information. For the $\vert \Psi^{+} \rangle$ and $\vert \mathrm{GHZ}_{3} \rangle$ states, we observe that setting the threshold to those identified by the Gini index captures the correct boundary where the minimal number of projectors is measured. Indeed, we observe that further reducing this value corresponds to a larger loss of information. Conversely, for the $\vert \mathrm{W}_{3} \rangle$ and $\vert \mathrm{GHZ}_{4} \rangle$ states we observe that setting the threshold to the one provided by the Gini index [Eq. (\ref{eq:threshold_gini})] corresponds to a slightly conservative choice, given that it is still possible to perform a further moderate reduction in the number of projectors without adding significant loss of information on the state. The same trend is confirmed by observing the purity of the reconstructed states with the tQST approach. For this parameter, the effect of removing an exceeding amount of projectors with respect to those indicated by the Gini index is found in a significant reduction of the reconstructed matrix purity, since not enough information was collected to properly estimate this parameter. As a final note, we observe that the tQST approach with threshold at the Gini index chooses a number of projectors which is different than the one expected for the corresponding noiseless states. Indeed, the presence of experimental noise adds other non-zero values in the diagonal elements, which then results in the need to measure additional projectors than those required for an ideal state.

%---------------------------------------------

\section{Conclusions and outlook}

In this manuscript, we have tested experimentally the application of the tQST approach in a hybrid photonic platform, verifying its application on states of up to $n=4$ qubits. The method is shown to provide an accurate reconstruction of unknown quantum states with different levels of sparsity in the density matrix, as testified by the fidelities achieved with the reconstructed states with respect to respect to those obtained by measuring a complete set of projectors. The advantage is found to be more pronounced for states having different elements with small (or zero) values in the density matrix, such as specific classes of maximally-entangled states which are at the basis of several quantum information protocols. According to the detected coincidence rates and considering that in our platform multiple projectors corresponding to the same basis are measured simultaneously in a single setting, application of the tQST method to the $n=4$ GHZ state led to a reduction in the time to measure the required projectors to approximately $\sim 7.5$ hrs with respect to $\sim 40.5$ hrs to implement tQST with threshold $t=0$. The obtained results thus show the effectiveness of the approach on a photonic platform, and opens new perspectives in its application in larger systems.

%---------------------------------------------

\section*{Acknowledgments}
This work is supported by PNRR MUR project PE0000023-NQSTI (Spoke 1, Spoke 4 and Spoke 7) and by the European Union’s Horizon Europe research and innovation program under EPIQUE Project (Grant Agreement No. 101135288). Fabrication of the femtosecond laser-written integrated circuit was partially performed at PoliFAB, the micro and nano-fabrication facility of Politecnico di Milano (https://www.polifab.polimi.it). F.C. and R.O. wish to thank the PoliFAB staff for the valuable technical support.

%---------------------------------------------

%\bibliographystyle{apsrev4-2}
%\bibliography{main.bib}

\begin{thebibliography}{43}%
\makeatletter
\providecommand \@ifxundefined [1]{%
 \@ifx{#1\undefined}
}%
\providecommand \@ifnum [1]{%
 \ifnum #1\expandafter \@firstoftwo
 \else \expandafter \@secondoftwo
 \fi
}%
\providecommand \@ifx [1]{%
 \ifx #1\expandafter \@firstoftwo
 \else \expandafter \@secondoftwo
 \fi
}%
\providecommand \natexlab [1]{#1}%
\providecommand \enquote  [1]{``#1''}%
\providecommand \bibnamefont  [1]{#1}%
\providecommand \bibfnamefont [1]{#1}%
\providecommand \citenamefont [1]{#1}%
\providecommand \href@noop [0]{\@secondoftwo}%
\providecommand \href [0]{\begingroup \@sanitize@url \@href}%
\providecommand \@href[1]{\@@startlink{#1}\@@href}%
\providecommand \@@href[1]{\endgroup#1\@@endlink}%
\providecommand \@sanitize@url [0]{\catcode `\\12\catcode `\$12\catcode
  `\&12\catcode `\#12\catcode `\^12\catcode `\_12\catcode `\%12\relax}%
\providecommand \@@startlink[1]{}%
\providecommand \@@endlink[0]{}%
\providecommand \url  [0]{\begingroup\@sanitize@url \@url }%
\providecommand \@url [1]{\endgroup\@href {#1}{\urlprefix }}%
\providecommand \urlprefix  [0]{URL }%
\providecommand \Eprint [0]{\href }%
\providecommand \doibase [0]{https://doi.org/}%
\providecommand \selectlanguage [0]{\@gobble}%
\providecommand \bibinfo  [0]{\@secondoftwo}%
\providecommand \bibfield  [0]{\@secondoftwo}%
\providecommand \translation [1]{[#1]}%
\providecommand \BibitemOpen [0]{}%
\providecommand \bibitemStop [0]{}%
\providecommand \bibitemNoStop [0]{.\EOS\space}%
\providecommand \EOS [0]{\spacefactor3000\relax}%
\providecommand \BibitemShut  [1]{\csname bibitem#1\endcsname}%
\let\auto@bib@innerbib\@empty
%</preamble>
\bibitem [{\citenamefont {James}\ \emph {et~al.}(2001)\citenamefont {James},
  \citenamefont {Kwiat}, \citenamefont {Munro},\ and\ \citenamefont
  {White}}]{james2001measurement}%
  \BibitemOpen
  \bibfield  {author} {\bibinfo {author} {\bibfnamefont {D.~F.}\ \bibnamefont
  {James}}, \bibinfo {author} {\bibfnamefont {P.~G.}\ \bibnamefont {Kwiat}},
  \bibinfo {author} {\bibfnamefont {W.~J.}\ \bibnamefont {Munro}},\ and\
  \bibinfo {author} {\bibfnamefont {A.~G.}\ \bibnamefont {White}},\ }\href
  {https://doi.org/10.1103/PhysRevA.64.052312} {\bibfield  {journal} {\bibinfo
  {journal} {Physical Review A}\ }\textbf {\bibinfo {volume} {64}},\ \bibinfo
  {pages} {052312} (\bibinfo {year} {2001})}\BibitemShut {NoStop}%
\bibitem [{\citenamefont {Thew}\ \emph {et~al.}(2002)\citenamefont {Thew},
  \citenamefont {Nemoto}, \citenamefont {White},\ and\ \citenamefont
  {Munro}}]{thew2002qudit}%
  \BibitemOpen
  \bibfield  {author} {\bibinfo {author} {\bibfnamefont {R.~T.}\ \bibnamefont
  {Thew}}, \bibinfo {author} {\bibfnamefont {K.}~\bibnamefont {Nemoto}},
  \bibinfo {author} {\bibfnamefont {A.~G.}\ \bibnamefont {White}},\ and\
  \bibinfo {author} {\bibfnamefont {W.~J.}\ \bibnamefont {Munro}},\ }\href
  {https://doi.org/10.1103/PhysRevA.66.012303} {\bibfield  {journal} {\bibinfo
  {journal} {Physical Review A}\ }\textbf {\bibinfo {volume} {66}},\ \bibinfo
  {pages} {012303} (\bibinfo {year} {2002})}\BibitemShut {NoStop}%
\bibitem [{\citenamefont {D'Ariano}\ \emph {et~al.}(2003)\citenamefont
  {D'Ariano}, \citenamefont {Paris},\ and\ \citenamefont
  {Sacchi}}]{d2003quantum}%
  \BibitemOpen
  \bibfield  {author} {\bibinfo {author} {\bibfnamefont {G.~M.}\ \bibnamefont
  {D'Ariano}}, \bibinfo {author} {\bibfnamefont {M.~G.}\ \bibnamefont
  {Paris}},\ and\ \bibinfo {author} {\bibfnamefont {M.~F.}\ \bibnamefont
  {Sacchi}},\ }\href {https://doi.org/10.1016/S1076-5670(03)80065-4} {\bibfield
   {journal} {\bibinfo  {journal} {Advances in imaging and electron physics}\
  }\textbf {\bibinfo {volume} {128}},\ \bibinfo {pages} {205} (\bibinfo {year}
  {2003})}\BibitemShut {NoStop}%
\bibitem [{\citenamefont {Altepeter}\ \emph {et~al.}(2005)\citenamefont
  {Altepeter}, \citenamefont {Jeffrey},\ and\ \citenamefont
  {Kwiat}}]{altepeter2005photonic}%
  \BibitemOpen
  \bibfield  {author} {\bibinfo {author} {\bibfnamefont {J.~B.}\ \bibnamefont
  {Altepeter}}, \bibinfo {author} {\bibfnamefont {E.~R.}\ \bibnamefont
  {Jeffrey}},\ and\ \bibinfo {author} {\bibfnamefont {P.~G.}\ \bibnamefont
  {Kwiat}},\ }\href {https://doi.org/10.1016/S1049-250X(05)52003-2} {\bibfield
  {journal} {\bibinfo  {journal} {Advances in atomic, molecular, and optical
  physics}\ }\textbf {\bibinfo {volume} {52}},\ \bibinfo {pages} {105}
  (\bibinfo {year} {2005})}\BibitemShut {NoStop}%
\bibitem [{\citenamefont {Gross}\ \emph {et~al.}(2010)\citenamefont {Gross},
  \citenamefont {Liu}, \citenamefont {Flammia}, \citenamefont {Becker},\ and\
  \citenamefont {Eisert}}]{gross2010quantum}%
  \BibitemOpen
  \bibfield  {author} {\bibinfo {author} {\bibfnamefont {D.}~\bibnamefont
  {Gross}}, \bibinfo {author} {\bibfnamefont {Y.-K.}\ \bibnamefont {Liu}},
  \bibinfo {author} {\bibfnamefont {S.~T.}\ \bibnamefont {Flammia}}, \bibinfo
  {author} {\bibfnamefont {S.}~\bibnamefont {Becker}},\ and\ \bibinfo {author}
  {\bibfnamefont {J.}~\bibnamefont {Eisert}},\ }\href
  {https://doi.org/10.1103/PhysRevLett.105.150401} {\bibfield  {journal}
  {\bibinfo  {journal} {Physical Review Letters}\ }\textbf {\bibinfo {volume}
  {105}},\ \bibinfo {pages} {150401} (\bibinfo {year} {2010})}\BibitemShut
  {NoStop}%
\bibitem [{\citenamefont {Flammia}\ \emph {et~al.}(2012)\citenamefont
  {Flammia}, \citenamefont {Gross}, \citenamefont {Liu},\ and\ \citenamefont
  {Eisert}}]{flammia2012quantum}%
  \BibitemOpen
  \bibfield  {author} {\bibinfo {author} {\bibfnamefont {S.~T.}\ \bibnamefont
  {Flammia}}, \bibinfo {author} {\bibfnamefont {D.}~\bibnamefont {Gross}},
  \bibinfo {author} {\bibfnamefont {Y.-K.}\ \bibnamefont {Liu}},\ and\ \bibinfo
  {author} {\bibfnamefont {J.}~\bibnamefont {Eisert}},\ }\href
  {https://doi.org/10.1088/1367-2630/14/9/095022} {\bibfield  {journal}
  {\bibinfo  {journal} {New Journal of Physics}\ }\textbf {\bibinfo {volume}
  {14}},\ \bibinfo {pages} {095022} (\bibinfo {year} {2012})}\BibitemShut
  {NoStop}%
\bibitem [{\citenamefont {Tonolini}\ \emph {et~al.}(2014)\citenamefont
  {Tonolini}, \citenamefont {Chan}, \citenamefont {Agnew}, \citenamefont
  {Lindsay},\ and\ \citenamefont {Leach}}]{tonolini2014reconstructing}%
  \BibitemOpen
  \bibfield  {author} {\bibinfo {author} {\bibfnamefont {F.}~\bibnamefont
  {Tonolini}}, \bibinfo {author} {\bibfnamefont {S.}~\bibnamefont {Chan}},
  \bibinfo {author} {\bibfnamefont {M.}~\bibnamefont {Agnew}}, \bibinfo
  {author} {\bibfnamefont {A.}~\bibnamefont {Lindsay}},\ and\ \bibinfo {author}
  {\bibfnamefont {J.}~\bibnamefont {Leach}},\ }\href
  {https://doi.org/10.1038/srep06542} {\bibfield  {journal} {\bibinfo
  {journal} {Scientific Reports}\ }\textbf {\bibinfo {volume} {4}},\ \bibinfo
  {pages} {6542} (\bibinfo {year} {2014})}\BibitemShut {NoStop}%
\bibitem [{\citenamefont {Blume-Kohout}(2010)}]{blume2010optimal}%
  \BibitemOpen
  \bibfield  {author} {\bibinfo {author} {\bibfnamefont {R.}~\bibnamefont
  {Blume-Kohout}},\ }\href {https://doi.org/10.1088/1367-2630/12/4/043034}
  {\bibfield  {journal} {\bibinfo  {journal} {New Journal of Physics}\ }\textbf
  {\bibinfo {volume} {12}},\ \bibinfo {pages} {043034} (\bibinfo {year}
  {2010})}\BibitemShut {NoStop}%
\bibitem [{\citenamefont {Lu}\ \emph {et~al.}(2018)\citenamefont {Lu},
  \citenamefont {Lukens}, \citenamefont {Peters}, \citenamefont {Williams},
  \citenamefont {Weiner},\ and\ \citenamefont {Lougovski}}]{lu2018quantum}%
  \BibitemOpen
  \bibfield  {author} {\bibinfo {author} {\bibfnamefont {H.-H.}\ \bibnamefont
  {Lu}}, \bibinfo {author} {\bibfnamefont {J.~M.}\ \bibnamefont {Lukens}},
  \bibinfo {author} {\bibfnamefont {N.~A.}\ \bibnamefont {Peters}}, \bibinfo
  {author} {\bibfnamefont {B.~P.}\ \bibnamefont {Williams}}, \bibinfo {author}
  {\bibfnamefont {A.~M.}\ \bibnamefont {Weiner}},\ and\ \bibinfo {author}
  {\bibfnamefont {P.}~\bibnamefont {Lougovski}},\ }\href
  {https://doi.org/10.1364/OPTICA.5.001455} {\bibfield  {journal} {\bibinfo
  {journal} {Optica}\ }\textbf {\bibinfo {volume} {5}},\ \bibinfo {pages}
  {1455} (\bibinfo {year} {2018})}\BibitemShut {NoStop}%
\bibitem [{\citenamefont {Lu}\ \emph {et~al.}(2022)\citenamefont {Lu},
  \citenamefont {Myilswamy}, \citenamefont {Bennink}, \citenamefont {Seshadri},
  \citenamefont {Alshaykh}, \citenamefont {Liu}, \citenamefont {Kippenberg},
  \citenamefont {Leaird}, \citenamefont {Weiner},\ and\ \citenamefont
  {Lukens}}]{lu2022bayesian}%
  \BibitemOpen
  \bibfield  {author} {\bibinfo {author} {\bibfnamefont {H.-H.}\ \bibnamefont
  {Lu}}, \bibinfo {author} {\bibfnamefont {K.~V.}\ \bibnamefont {Myilswamy}},
  \bibinfo {author} {\bibfnamefont {R.~S.}\ \bibnamefont {Bennink}}, \bibinfo
  {author} {\bibfnamefont {S.}~\bibnamefont {Seshadri}}, \bibinfo {author}
  {\bibfnamefont {M.~S.}\ \bibnamefont {Alshaykh}}, \bibinfo {author}
  {\bibfnamefont {J.}~\bibnamefont {Liu}}, \bibinfo {author} {\bibfnamefont
  {T.~J.}\ \bibnamefont {Kippenberg}}, \bibinfo {author} {\bibfnamefont
  {D.~E.}\ \bibnamefont {Leaird}}, \bibinfo {author} {\bibfnamefont {A.~M.}\
  \bibnamefont {Weiner}},\ and\ \bibinfo {author} {\bibfnamefont {J.~M.}\
  \bibnamefont {Lukens}},\ }\href {https://doi.org/10.1038/s41467-022-31639-z}
  {\bibfield  {journal} {\bibinfo  {journal} {Nature Communications}\ }\textbf
  {\bibinfo {volume} {13}},\ \bibinfo {pages} {4338} (\bibinfo {year}
  {2022})}\BibitemShut {NoStop}%
\bibitem [{\citenamefont {Aaronson}(2018)}]{aaronson2018shadow}%
  \BibitemOpen
  \bibfield  {author} {\bibinfo {author} {\bibfnamefont {S.}~\bibnamefont
  {Aaronson}},\ }in\ \href {https://doi.org/10.1145/3188745.31888} {\emph
  {\bibinfo {booktitle} {Proceedings of the 50th annual ACM SIGACT symposium on
  theory of computing}}}\ (\bibinfo  {publisher} {{ACM} Press},\ \bibinfo
  {year} {2018})\ pp.\ \bibinfo {pages} {325--338}\BibitemShut {NoStop}%
\bibitem [{\citenamefont {Huang}\ \emph {et~al.}(2020)\citenamefont {Huang},
  \citenamefont {Kueng},\ and\ \citenamefont {Preskill}}]{huang2020predicting}%
  \BibitemOpen
  \bibfield  {author} {\bibinfo {author} {\bibfnamefont {H.-Y.}\ \bibnamefont
  {Huang}}, \bibinfo {author} {\bibfnamefont {R.}~\bibnamefont {Kueng}},\ and\
  \bibinfo {author} {\bibfnamefont {J.}~\bibnamefont {Preskill}},\ }\href
  {https://doi.org/10.1038/s41567-020-0932-7} {\bibfield  {journal} {\bibinfo
  {journal} {Nature Physics}\ }\textbf {\bibinfo {volume} {16}},\ \bibinfo
  {pages} {1050} (\bibinfo {year} {2020})}\BibitemShut {NoStop}%
\bibitem [{\citenamefont {Struchalin}\ \emph {et~al.}(2021)\citenamefont
  {Struchalin}, \citenamefont {Zagorovskii}, \citenamefont {Kovlakov},
  \citenamefont {Straupe},\ and\ \citenamefont
  {Kulik}}]{struchalin2021experimental}%
  \BibitemOpen
  \bibfield  {author} {\bibinfo {author} {\bibfnamefont {G.}~\bibnamefont
  {Struchalin}}, \bibinfo {author} {\bibfnamefont {Y.~A.}\ \bibnamefont
  {Zagorovskii}}, \bibinfo {author} {\bibfnamefont {E.}~\bibnamefont
  {Kovlakov}}, \bibinfo {author} {\bibfnamefont {S.}~\bibnamefont {Straupe}},\
  and\ \bibinfo {author} {\bibfnamefont {S.}~\bibnamefont {Kulik}},\ }\href
  {https://doi.org/10.1103/PRXQuantum.2.010307} {\bibfield  {journal} {\bibinfo
   {journal} {PRX Quantum}\ }\textbf {\bibinfo {volume} {2}},\ \bibinfo {pages}
  {010307} (\bibinfo {year} {2021})}\BibitemShut {NoStop}%
\bibitem [{\citenamefont {Binosi}\ \emph {et~al.}(2024)\citenamefont {Binosi},
  \citenamefont {Garberoglio}, \citenamefont {Maragnano}, \citenamefont
  {Dapor},\ and\ \citenamefont {Liscidini}}]{binosi2024tailor}%
  \BibitemOpen
  \bibfield  {author} {\bibinfo {author} {\bibfnamefont {D.}~\bibnamefont
  {Binosi}}, \bibinfo {author} {\bibfnamefont {G.}~\bibnamefont {Garberoglio}},
  \bibinfo {author} {\bibfnamefont {D.}~\bibnamefont {Maragnano}}, \bibinfo
  {author} {\bibfnamefont {M.}~\bibnamefont {Dapor}},\ and\ \bibinfo {author}
  {\bibfnamefont {M.}~\bibnamefont {Liscidini}},\ }\href
  {https://doi.org/10.1063/5.0219143} {\bibfield  {journal} {\bibinfo
  {journal} {APL Quantum}\ }\textbf {\bibinfo {volume} {1}},\ \bibinfo {pages}
  {036112} (\bibinfo {year} {2024})}\BibitemShut {NoStop}%
\bibitem [{\citenamefont {Wang}\ \emph {et~al.}(2020)\citenamefont {Wang},
  \citenamefont {Sciarrino}, \citenamefont {Laing},\ and\ \citenamefont
  {Thompson}}]{wang2020integrated}%
  \BibitemOpen
  \bibfield  {author} {\bibinfo {author} {\bibfnamefont {J.}~\bibnamefont
  {Wang}}, \bibinfo {author} {\bibfnamefont {F.}~\bibnamefont {Sciarrino}},
  \bibinfo {author} {\bibfnamefont {A.}~\bibnamefont {Laing}},\ and\ \bibinfo
  {author} {\bibfnamefont {M.~G.}\ \bibnamefont {Thompson}},\ }\href
  {https://doi.org/10.1038/s41566-019-0532-1} {\bibfield  {journal} {\bibinfo
  {journal} {Nature Photonics}\ }\textbf {\bibinfo {volume} {14}},\ \bibinfo
  {pages} {273} (\bibinfo {year} {2020})}\BibitemShut {NoStop}%
\bibitem [{\citenamefont {Giordani}\ \emph {et~al.}(2023)\citenamefont
  {Giordani}, \citenamefont {Hoch}, \citenamefont {Carvacho}, \citenamefont
  {Spagnolo},\ and\ \citenamefont {Sciarrino}}]{giordani2023integrated}%
  \BibitemOpen
  \bibfield  {author} {\bibinfo {author} {\bibfnamefont {T.}~\bibnamefont
  {Giordani}}, \bibinfo {author} {\bibfnamefont {F.}~\bibnamefont {Hoch}},
  \bibinfo {author} {\bibfnamefont {G.}~\bibnamefont {Carvacho}}, \bibinfo
  {author} {\bibfnamefont {N.}~\bibnamefont {Spagnolo}},\ and\ \bibinfo
  {author} {\bibfnamefont {F.}~\bibnamefont {Sciarrino}},\ }\href
  {https://doi.org/10.1007/s40766-023-00040-x} {\bibfield  {journal} {\bibinfo
  {journal} {La Rivista del Nuovo Cimento}\ }\textbf {\bibinfo {volume} {46}},\
  \bibinfo {pages} {71} (\bibinfo {year} {2023})}\BibitemShut {NoStop}%
\bibitem [{\citenamefont {Ahn}\ \emph {et~al.}(2019{\natexlab{a}})\citenamefont
  {Ahn}, \citenamefont {Teo}, \citenamefont {Jeong}, \citenamefont {Bouchard},
  \citenamefont {Hufnagel}, \citenamefont {Karimi}, \citenamefont {Koutn{\`y}},
  \citenamefont {{\v{R}}eh{\'a}{\v{c}}ek}, \citenamefont {Hradil},
  \citenamefont {Leuchs},\ and\ \citenamefont
  {S\'anchez-Soto}}]{ahn2019adaptive}%
  \BibitemOpen
  \bibfield  {author} {\bibinfo {author} {\bibfnamefont {D.}~\bibnamefont
  {Ahn}}, \bibinfo {author} {\bibfnamefont {Y.~S.}\ \bibnamefont {Teo}},
  \bibinfo {author} {\bibfnamefont {H.}~\bibnamefont {Jeong}}, \bibinfo
  {author} {\bibfnamefont {F.}~\bibnamefont {Bouchard}}, \bibinfo {author}
  {\bibfnamefont {F.}~\bibnamefont {Hufnagel}}, \bibinfo {author}
  {\bibfnamefont {E.}~\bibnamefont {Karimi}}, \bibinfo {author} {\bibfnamefont
  {D.}~\bibnamefont {Koutn{\`y}}}, \bibinfo {author} {\bibfnamefont
  {J.}~\bibnamefont {{\v{R}}eh{\'a}{\v{c}}ek}}, \bibinfo {author}
  {\bibfnamefont {Z.}~\bibnamefont {Hradil}}, \bibinfo {author} {\bibfnamefont
  {G.}~\bibnamefont {Leuchs}},\ and\ \bibinfo {author} {\bibfnamefont {L.~L.}\
  \bibnamefont {S\'anchez-Soto}},\ }\href
  {https://doi.org/10.1103/PhysRevLett.122.100404} {\bibfield  {journal}
  {\bibinfo  {journal} {Physical Review Letters}\ }\textbf {\bibinfo {volume}
  {122}},\ \bibinfo {pages} {100404} (\bibinfo {year}
  {2019}{\natexlab{a}})}\BibitemShut {NoStop}%
\bibitem [{\citenamefont {Ahn}\ \emph {et~al.}(2019{\natexlab{b}})\citenamefont
  {Ahn}, \citenamefont {Teo}, \citenamefont {Jeong}, \citenamefont
  {Koutn{\`y}}, \citenamefont {{\v{R}}eh{\'a}{\v{c}}ek}, \citenamefont
  {Hradil}, \citenamefont {Leuchs},\ and\ \citenamefont
  {S{\'a}nchez-Soto}}]{ahn2019adaptive_numerical}%
  \BibitemOpen
  \bibfield  {author} {\bibinfo {author} {\bibfnamefont {D.}~\bibnamefont
  {Ahn}}, \bibinfo {author} {\bibfnamefont {Y.}~\bibnamefont {Teo}}, \bibinfo
  {author} {\bibfnamefont {H.}~\bibnamefont {Jeong}}, \bibinfo {author}
  {\bibfnamefont {D.}~\bibnamefont {Koutn{\`y}}}, \bibinfo {author}
  {\bibfnamefont {J.}~\bibnamefont {{\v{R}}eh{\'a}{\v{c}}ek}}, \bibinfo
  {author} {\bibfnamefont {Z.}~\bibnamefont {Hradil}}, \bibinfo {author}
  {\bibfnamefont {G.}~\bibnamefont {Leuchs}},\ and\ \bibinfo {author}
  {\bibfnamefont {L.}~\bibnamefont {S{\'a}nchez-Soto}},\ }\href
  {https://doi.org/10.1103/PhysRevA.100.012346} {\bibfield  {journal} {\bibinfo
   {journal} {Physical Review A}\ }\textbf {\bibinfo {volume} {100}},\ \bibinfo
  {pages} {012346} (\bibinfo {year} {2019}{\natexlab{b}})}\BibitemShut
  {NoStop}%
\bibitem [{\citenamefont {Baumgratz}\ \emph {et~al.}(2013)\citenamefont
  {Baumgratz}, \citenamefont {Gross}, \citenamefont {Cramer},\ and\
  \citenamefont {Plenio}}]{baumgratz2013scalable}%
  \BibitemOpen
  \bibfield  {author} {\bibinfo {author} {\bibfnamefont {T.}~\bibnamefont
  {Baumgratz}}, \bibinfo {author} {\bibfnamefont {D.}~\bibnamefont {Gross}},
  \bibinfo {author} {\bibfnamefont {M.}~\bibnamefont {Cramer}},\ and\ \bibinfo
  {author} {\bibfnamefont {M.~B.}\ \bibnamefont {Plenio}},\ }\href
  {https://doi.org/10.1103/PhysRevLett.111.020401} {\bibfield  {journal}
  {\bibinfo  {journal} {Physical review letters}\ }\textbf {\bibinfo {volume}
  {111}},\ \bibinfo {pages} {020401} (\bibinfo {year} {2013})}\BibitemShut
  {NoStop}%
\bibitem [{\citenamefont {Hurley}\ and\ \citenamefont
  {Rickard}(2009)}]{hurley2009comparing}%
  \BibitemOpen
  \bibfield  {author} {\bibinfo {author} {\bibfnamefont {N.}~\bibnamefont
  {Hurley}}\ and\ \bibinfo {author} {\bibfnamefont {S.}~\bibnamefont
  {Rickard}},\ }\href {https://doi.org/10.1109/TIT.2009.2027527} {\bibfield
  {journal} {\bibinfo  {journal} {IEEE Transactions on Information Theory}\
  }\textbf {\bibinfo {volume} {55}},\ \bibinfo {pages} {4723} (\bibinfo {year}
  {2009})}\BibitemShut {NoStop}%
\bibitem [{\citenamefont {Elshaari}\ \emph {et~al.}(2020)\citenamefont
  {Elshaari}, \citenamefont {Pernice}, \citenamefont {Srinivasan},
  \citenamefont {Benson},\ and\ \citenamefont {Zwiller}}]{elshaari2020hybrid}%
  \BibitemOpen
  \bibfield  {author} {\bibinfo {author} {\bibfnamefont {A.~W.}\ \bibnamefont
  {Elshaari}}, \bibinfo {author} {\bibfnamefont {W.}~\bibnamefont {Pernice}},
  \bibinfo {author} {\bibfnamefont {K.}~\bibnamefont {Srinivasan}}, \bibinfo
  {author} {\bibfnamefont {O.}~\bibnamefont {Benson}},\ and\ \bibinfo {author}
  {\bibfnamefont {V.}~\bibnamefont {Zwiller}},\ }\href
  {https://doi.org/10.1038/s41566-020-0609-x} {\bibfield  {journal} {\bibinfo
  {journal} {Nature photonics}\ }\textbf {\bibinfo {volume} {14}},\ \bibinfo
  {pages} {285} (\bibinfo {year} {2020})}\BibitemShut {NoStop}%
\bibitem [{\citenamefont {Rodari}\ \emph
  {et~al.}(2024{\natexlab{a}})\citenamefont {Rodari}, \citenamefont {Novo},
  \citenamefont {Albiero}, \citenamefont {Suprano}, \citenamefont {Tavares},
  \citenamefont {Caruccio}, \citenamefont {Hoch}, \citenamefont {Giordani},
  \citenamefont {Carvacho}, \citenamefont {Gardina}, \citenamefont {Di~Giano},
  \citenamefont {Di~Giorgio}, \citenamefont {Corrielli}, \citenamefont
  {Ceccarelli}, \citenamefont {Osellame}, \citenamefont {Spagnolo},
  \citenamefont {Galv\"{a}o},\ and\ \citenamefont
  {Sciarrino}}]{rodari2024semi}%
  \BibitemOpen
  \bibfield  {author} {\bibinfo {author} {\bibfnamefont {G.}~\bibnamefont
  {Rodari}}, \bibinfo {author} {\bibfnamefont {L.}~\bibnamefont {Novo}},
  \bibinfo {author} {\bibfnamefont {R.}~\bibnamefont {Albiero}}, \bibinfo
  {author} {\bibfnamefont {A.}~\bibnamefont {Suprano}}, \bibinfo {author}
  {\bibfnamefont {C.~T.}\ \bibnamefont {Tavares}}, \bibinfo {author}
  {\bibfnamefont {E.}~\bibnamefont {Caruccio}}, \bibinfo {author}
  {\bibfnamefont {F.}~\bibnamefont {Hoch}}, \bibinfo {author} {\bibfnamefont
  {T.}~\bibnamefont {Giordani}}, \bibinfo {author} {\bibfnamefont
  {G.}~\bibnamefont {Carvacho}}, \bibinfo {author} {\bibfnamefont
  {M.}~\bibnamefont {Gardina}}, \bibinfo {author} {\bibfnamefont
  {N.}~\bibnamefont {Di~Giano}}, \bibinfo {author} {\bibfnamefont
  {S.}~\bibnamefont {Di~Giorgio}}, \bibinfo {author} {\bibfnamefont
  {G.}~\bibnamefont {Corrielli}}, \bibinfo {author} {\bibfnamefont
  {F.}~\bibnamefont {Ceccarelli}}, \bibinfo {author} {\bibfnamefont
  {R.}~\bibnamefont {Osellame}}, \bibinfo {author} {\bibfnamefont
  {N.}~\bibnamefont {Spagnolo}}, \bibinfo {author} {\bibfnamefont {E.~F.}\
  \bibnamefont {Galv\"{a}o}},\ and\ \bibinfo {author} {\bibfnamefont
  {F.}~\bibnamefont {Sciarrino}},\ }\href@noop {} {\bibinfo {title}
  {Semi-device independent characterization of multiphoton
  indistinguishability}} (\bibinfo {year} {2024}{\natexlab{a}}),\ \Eprint
  {https://arxiv.org/abs/2404.18636} {arXiv:2404.18636 [quant-ph]} \BibitemShut
  {NoStop}%
\bibitem [{\citenamefont {Rodari}\ \emph
  {et~al.}(2024{\natexlab{b}})\citenamefont {Rodari}, \citenamefont
  {Fernandes}, \citenamefont {Caruccio}, \citenamefont {Suprano}, \citenamefont
  {Hoch}, \citenamefont {Giordani}, \citenamefont {Carvacho}, \citenamefont
  {Albiero}, \citenamefont {Di~Giano}, \citenamefont {Corrielli}, \citenamefont
  {Ceccarelli}, \citenamefont {Osellame}, \citenamefont {Brod}, \citenamefont
  {Novo}, \citenamefont {Spagnolo}, \citenamefont {Galv\"{a}o},\ and\
  \citenamefont {Sciarrino}}]{rodari2024experimental}%
  \BibitemOpen
  \bibfield  {author} {\bibinfo {author} {\bibfnamefont {G.}~\bibnamefont
  {Rodari}}, \bibinfo {author} {\bibfnamefont {C.}~\bibnamefont {Fernandes}},
  \bibinfo {author} {\bibfnamefont {E.}~\bibnamefont {Caruccio}}, \bibinfo
  {author} {\bibfnamefont {A.}~\bibnamefont {Suprano}}, \bibinfo {author}
  {\bibfnamefont {F.}~\bibnamefont {Hoch}}, \bibinfo {author} {\bibfnamefont
  {T.}~\bibnamefont {Giordani}}, \bibinfo {author} {\bibfnamefont
  {G.}~\bibnamefont {Carvacho}}, \bibinfo {author} {\bibfnamefont
  {R.}~\bibnamefont {Albiero}}, \bibinfo {author} {\bibfnamefont
  {N.}~\bibnamefont {Di~Giano}}, \bibinfo {author} {\bibfnamefont
  {G.}~\bibnamefont {Corrielli}}, \bibinfo {author} {\bibfnamefont
  {F.}~\bibnamefont {Ceccarelli}}, \bibinfo {author} {\bibfnamefont
  {R.}~\bibnamefont {Osellame}}, \bibinfo {author} {\bibfnamefont {D.~J.}\
  \bibnamefont {Brod}}, \bibinfo {author} {\bibfnamefont {L.}~\bibnamefont
  {Novo}}, \bibinfo {author} {\bibfnamefont {N.}~\bibnamefont {Spagnolo}},
  \bibinfo {author} {\bibfnamefont {E.~F.}\ \bibnamefont {Galv\"{a}o}},\ and\
  \bibinfo {author} {\bibfnamefont {F.}~\bibnamefont {Sciarrino}},\ }\href@noop
  {} {\bibinfo {title} {Experimental observation of counter-intuitive features
  of photonic bunching}} (\bibinfo {year} {2024}{\natexlab{b}}),\ \Eprint
  {https://arxiv.org/abs/2410.15883} {arXiv:2410.15883 [quant-ph]} \BibitemShut
  {NoStop}%
\bibitem [{\citenamefont {Michler}(2017)}]{michler2017quantum}%
  \BibitemOpen
  \bibfield  {author} {\bibinfo {author} {\bibfnamefont {P.}~\bibnamefont
  {Michler}},\ }\href {https://doi.org/10.1007/978-3-319-56378-7} {\emph
  {\bibinfo {title} {Quantum dots for quantum information technologies}}},\
  Vol.\ \bibinfo {volume} {237}\ (\bibinfo  {publisher} {Springer},\ \bibinfo
  {year} {2017})\BibitemShut {NoStop}%
\bibitem [{\citenamefont {Senellart}\ \emph {et~al.}(2017)\citenamefont
  {Senellart}, \citenamefont {Solomon},\ and\ \citenamefont
  {White}}]{senellart2017high}%
  \BibitemOpen
  \bibfield  {author} {\bibinfo {author} {\bibfnamefont {P.}~\bibnamefont
  {Senellart}}, \bibinfo {author} {\bibfnamefont {G.}~\bibnamefont {Solomon}},\
  and\ \bibinfo {author} {\bibfnamefont {A.}~\bibnamefont {White}},\ }\href
  {https://doi.org/10.1038/nnano.2017.218} {\bibfield  {journal} {\bibinfo
  {journal} {Nature Nanotechnology}\ }\textbf {\bibinfo {volume} {12}},\
  \bibinfo {pages} {1026} (\bibinfo {year} {2017})}\BibitemShut {NoStop}%
\bibitem [{\citenamefont {Heindel}\ \emph {et~al.}(2023)\citenamefont
  {Heindel}, \citenamefont {Kim}, \citenamefont {Gregersen}, \citenamefont
  {Rastelli},\ and\ \citenamefont {Reitzenstein}}]{heindel2023quantum}%
  \BibitemOpen
  \bibfield  {author} {\bibinfo {author} {\bibfnamefont {T.}~\bibnamefont
  {Heindel}}, \bibinfo {author} {\bibfnamefont {J.-H.}\ \bibnamefont {Kim}},
  \bibinfo {author} {\bibfnamefont {N.}~\bibnamefont {Gregersen}}, \bibinfo
  {author} {\bibfnamefont {A.}~\bibnamefont {Rastelli}},\ and\ \bibinfo
  {author} {\bibfnamefont {S.}~\bibnamefont {Reitzenstein}},\ }\href
  {https://doi.org/10.1364/AOP.490091} {\bibfield  {journal} {\bibinfo
  {journal} {Advances in Optics and Photonics}\ }\textbf {\bibinfo {volume}
  {15}},\ \bibinfo {pages} {613} (\bibinfo {year} {2023})}\BibitemShut
  {NoStop}%
\bibitem [{\citenamefont {Esmann}\ \emph {et~al.}(2024)\citenamefont {Esmann},
  \citenamefont {Wein},\ and\ \citenamefont
  {Ant{\'o}n-Solanas}}]{esmann2024solid}%
  \BibitemOpen
  \bibfield  {author} {\bibinfo {author} {\bibfnamefont {M.}~\bibnamefont
  {Esmann}}, \bibinfo {author} {\bibfnamefont {S.~C.}\ \bibnamefont {Wein}},\
  and\ \bibinfo {author} {\bibfnamefont {C.}~\bibnamefont
  {Ant{\'o}n-Solanas}},\ }\href {https://doi.org/10.1002/adfm.202315936}
  {\bibfield  {journal} {\bibinfo  {journal} {Advanced Functional Materials}\
  }\textbf {\bibinfo {volume} {34}},\ \bibinfo {pages} {2315936} (\bibinfo
  {year} {2024})}\BibitemShut {NoStop}%
\bibitem [{\citenamefont {Somaschi}\ \emph {et~al.}(2016)\citenamefont
  {Somaschi}, \citenamefont {Giesz}, \citenamefont {De~Santis}, \citenamefont
  {Loredo}, \citenamefont {Almeida}, \citenamefont {Hornecker}, \citenamefont
  {Portalupi}, \citenamefont {Grange}, \citenamefont {Anton}, \citenamefont
  {Demory}, \citenamefont {G\'omez}, \citenamefont {Sagnes}, \citenamefont
  {Lanzillotti-Kimura}, \citenamefont {Lema{\^\i}tre}, \citenamefont
  {Auffeves}, \citenamefont {White},\ and\ \citenamefont
  {Lanco}}]{somaschi2016near}%
  \BibitemOpen
  \bibfield  {author} {\bibinfo {author} {\bibfnamefont {N.}~\bibnamefont
  {Somaschi}}, \bibinfo {author} {\bibfnamefont {V.}~\bibnamefont {Giesz}},
  \bibinfo {author} {\bibfnamefont {L.}~\bibnamefont {De~Santis}}, \bibinfo
  {author} {\bibfnamefont {J.}~\bibnamefont {Loredo}}, \bibinfo {author}
  {\bibfnamefont {M.~P.}\ \bibnamefont {Almeida}}, \bibinfo {author}
  {\bibfnamefont {G.}~\bibnamefont {Hornecker}}, \bibinfo {author}
  {\bibfnamefont {S.~L.}\ \bibnamefont {Portalupi}}, \bibinfo {author}
  {\bibfnamefont {T.}~\bibnamefont {Grange}}, \bibinfo {author} {\bibfnamefont
  {C.}~\bibnamefont {Anton}}, \bibinfo {author} {\bibfnamefont
  {J.}~\bibnamefont {Demory}}, \bibinfo {author} {\bibfnamefont
  {L.}~\bibnamefont {G\'omez}}, \bibinfo {author} {\bibfnamefont
  {I.}~\bibnamefont {Sagnes}}, \bibinfo {author} {\bibfnamefont {N.~D.}\
  \bibnamefont {Lanzillotti-Kimura}}, \bibinfo {author} {\bibfnamefont
  {A.}~\bibnamefont {Lema{\^\i}tre}}, \bibinfo {author} {\bibfnamefont
  {A.}~\bibnamefont {Auffeves}}, \bibinfo {author} {\bibfnamefont {A.~G.}\
  \bibnamefont {White}},\ and\ \bibinfo {author} {\bibfnamefont
  {P.}~\bibnamefont {Lanco}, \bibfnamefont {L~an~Senellart}},\ }\href
  {https://doi.org/10.1038/nphoton.2016.23} {\bibfield  {journal} {\bibinfo
  {journal} {Nature Photonics}\ }\textbf {\bibinfo {volume} {10}},\ \bibinfo
  {pages} {340} (\bibinfo {year} {2016})}\BibitemShut {NoStop}%
\bibitem [{\citenamefont {Ollivier}\ \emph {et~al.}(2020)\citenamefont
  {Ollivier}, \citenamefont {Maillette~de Buy~Wenniger}, \citenamefont
  {Thomas}, \citenamefont {Wein}, \citenamefont {Harouri}, \citenamefont
  {Coppola}, \citenamefont {Hilaire}, \citenamefont {Millet}, \citenamefont
  {Lema{\^\i}tre}, \citenamefont {Sagnes}, \citenamefont {Krebs}, \citenamefont
  {Lanco}, \citenamefont {Loredo}, \citenamefont {Ant\'on}, \citenamefont
  {Somaschi},\ and\ \citenamefont {Senellart}}]{ollivier2020reproducibility}%
  \BibitemOpen
  \bibfield  {author} {\bibinfo {author} {\bibfnamefont {H.}~\bibnamefont
  {Ollivier}}, \bibinfo {author} {\bibfnamefont {I.}~\bibnamefont {Maillette~de
  Buy~Wenniger}}, \bibinfo {author} {\bibfnamefont {S.}~\bibnamefont {Thomas}},
  \bibinfo {author} {\bibfnamefont {S.~C.}\ \bibnamefont {Wein}}, \bibinfo
  {author} {\bibfnamefont {A.}~\bibnamefont {Harouri}}, \bibinfo {author}
  {\bibfnamefont {G.}~\bibnamefont {Coppola}}, \bibinfo {author} {\bibfnamefont
  {P.}~\bibnamefont {Hilaire}}, \bibinfo {author} {\bibfnamefont
  {C.}~\bibnamefont {Millet}}, \bibinfo {author} {\bibfnamefont
  {A.}~\bibnamefont {Lema{\^\i}tre}}, \bibinfo {author} {\bibfnamefont
  {I.}~\bibnamefont {Sagnes}}, \bibinfo {author} {\bibfnamefont
  {O.}~\bibnamefont {Krebs}}, \bibinfo {author} {\bibfnamefont
  {L.}~\bibnamefont {Lanco}}, \bibinfo {author} {\bibfnamefont {J.~C.}\
  \bibnamefont {Loredo}}, \bibinfo {author} {\bibfnamefont {C.}~\bibnamefont
  {Ant\'on}}, \bibinfo {author} {\bibfnamefont {N.}~\bibnamefont {Somaschi}},\
  and\ \bibinfo {author} {\bibfnamefont {P.}~\bibnamefont {Senellart}},\ }\href
  {https://doi.org/10.1021/acsphotonics.9b01805} {\bibfield  {journal}
  {\bibinfo  {journal} {ACS Photonics}\ }\textbf {\bibinfo {volume} {7}},\
  \bibinfo {pages} {1050} (\bibinfo {year} {2020})}\BibitemShut {NoStop}%
\bibitem [{\citenamefont {Nowak}\ \emph {et~al.}(2014)\citenamefont {Nowak},
  \citenamefont {Portalupi}, \citenamefont {Giesz}, \citenamefont {Gazzano},
  \citenamefont {Dal~Savio}, \citenamefont {Braun}, \citenamefont {Karrai},
  \citenamefont {Arnold}, \citenamefont {Lanco}, \citenamefont {Sagnes},
  \citenamefont {Lema{\^\i}tre},\ and\ \citenamefont
  {Senellart}}]{nowak2014deterministic}%
  \BibitemOpen
  \bibfield  {author} {\bibinfo {author} {\bibfnamefont {A.}~\bibnamefont
  {Nowak}}, \bibinfo {author} {\bibfnamefont {S.}~\bibnamefont {Portalupi}},
  \bibinfo {author} {\bibfnamefont {V.}~\bibnamefont {Giesz}}, \bibinfo
  {author} {\bibfnamefont {O.}~\bibnamefont {Gazzano}}, \bibinfo {author}
  {\bibfnamefont {C.}~\bibnamefont {Dal~Savio}}, \bibinfo {author}
  {\bibfnamefont {P.-F.}\ \bibnamefont {Braun}}, \bibinfo {author}
  {\bibfnamefont {K.}~\bibnamefont {Karrai}}, \bibinfo {author} {\bibfnamefont
  {C.}~\bibnamefont {Arnold}}, \bibinfo {author} {\bibfnamefont
  {L.}~\bibnamefont {Lanco}}, \bibinfo {author} {\bibfnamefont
  {I.}~\bibnamefont {Sagnes}}, \bibinfo {author} {\bibfnamefont
  {A.}~\bibnamefont {Lema{\^\i}tre}},\ and\ \bibinfo {author} {\bibfnamefont
  {P.}~\bibnamefont {Senellart}},\ }\href {https://doi.org/10.1038/ncomms4240}
  {\bibfield  {journal} {\bibinfo  {journal} {Nature Communications}\ }\textbf
  {\bibinfo {volume} {5}},\ \bibinfo {pages} {3240} (\bibinfo {year}
  {2014})}\BibitemShut {NoStop}%
\bibitem [{\citenamefont {Gazzano}\ \emph {et~al.}(2013)\citenamefont
  {Gazzano}, \citenamefont {Michaelis~de Vasconcellos}, \citenamefont {Arnold},
  \citenamefont {Nowak}, \citenamefont {Galopin}, \citenamefont {Sagnes},
  \citenamefont {Lanco}, \citenamefont {Lema{\^\i}tre},\ and\ \citenamefont
  {Senellart}}]{gazzano2013bright}%
  \BibitemOpen
  \bibfield  {author} {\bibinfo {author} {\bibfnamefont {O.}~\bibnamefont
  {Gazzano}}, \bibinfo {author} {\bibfnamefont {S.}~\bibnamefont {Michaelis~de
  Vasconcellos}}, \bibinfo {author} {\bibfnamefont {C.}~\bibnamefont {Arnold}},
  \bibinfo {author} {\bibfnamefont {A.}~\bibnamefont {Nowak}}, \bibinfo
  {author} {\bibfnamefont {E.}~\bibnamefont {Galopin}}, \bibinfo {author}
  {\bibfnamefont {I.}~\bibnamefont {Sagnes}}, \bibinfo {author} {\bibfnamefont
  {L.}~\bibnamefont {Lanco}}, \bibinfo {author} {\bibfnamefont
  {A.}~\bibnamefont {Lema{\^\i}tre}},\ and\ \bibinfo {author} {\bibfnamefont
  {P.}~\bibnamefont {Senellart}},\ }\href {https://doi.org/10.1038/ncomms2434}
  {\bibfield  {journal} {\bibinfo  {journal} {Nature Communications}\ }\textbf
  {\bibinfo {volume} {4}},\ \bibinfo {pages} {1425} (\bibinfo {year}
  {2013})}\BibitemShut {NoStop}%
\bibitem [{\citenamefont {Hansen}\ \emph {et~al.}(2023)\citenamefont {Hansen},
  \citenamefont {Carosini}, \citenamefont {Jehle}, \citenamefont {Giorgino},
  \citenamefont {Houvenaghel}, \citenamefont {Vyvlecka}, \citenamefont
  {Loredo},\ and\ \citenamefont {Walther}}]{hansen2023single}%
  \BibitemOpen
  \bibfield  {author} {\bibinfo {author} {\bibfnamefont {L.~M.}\ \bibnamefont
  {Hansen}}, \bibinfo {author} {\bibfnamefont {L.}~\bibnamefont {Carosini}},
  \bibinfo {author} {\bibfnamefont {L.}~\bibnamefont {Jehle}}, \bibinfo
  {author} {\bibfnamefont {F.}~\bibnamefont {Giorgino}}, \bibinfo {author}
  {\bibfnamefont {R.}~\bibnamefont {Houvenaghel}}, \bibinfo {author}
  {\bibfnamefont {M.}~\bibnamefont {Vyvlecka}}, \bibinfo {author}
  {\bibfnamefont {J.~C.}\ \bibnamefont {Loredo}},\ and\ \bibinfo {author}
  {\bibfnamefont {P.}~\bibnamefont {Walther}},\ }\href
  {https://doi.org/10.1364/OPTICAQ.494643} {\bibfield  {journal} {\bibinfo
  {journal} {Optica Quantum}\ }\textbf {\bibinfo {volume} {1}},\ \bibinfo
  {pages} {1} (\bibinfo {year} {2023})}\BibitemShut {NoStop}%
\bibitem [{\citenamefont {Pont}\ \emph {et~al.}(2022)\citenamefont {Pont},
  \citenamefont {Albiero}, \citenamefont {Thomas}, \citenamefont {Spagnolo},
  \citenamefont {Ceccarelli}, \citenamefont {Corrielli}, \citenamefont
  {Brieussel}, \citenamefont {Somaschi}, \citenamefont {Huet}, \citenamefont
  {Harouri}, \citenamefont {Lema{\^\i}tre}, \citenamefont {Sagnes},
  \citenamefont {Belabas}, \citenamefont {Sciarrino}, \citenamefont {Osellame},
  \citenamefont {Senellart},\ and\ \citenamefont
  {Crespi}}]{pont2022quantifying}%
  \BibitemOpen
  \bibfield  {author} {\bibinfo {author} {\bibfnamefont {M.}~\bibnamefont
  {Pont}}, \bibinfo {author} {\bibfnamefont {R.}~\bibnamefont {Albiero}},
  \bibinfo {author} {\bibfnamefont {S.~E.}\ \bibnamefont {Thomas}}, \bibinfo
  {author} {\bibfnamefont {N.}~\bibnamefont {Spagnolo}}, \bibinfo {author}
  {\bibfnamefont {F.}~\bibnamefont {Ceccarelli}}, \bibinfo {author}
  {\bibfnamefont {G.}~\bibnamefont {Corrielli}}, \bibinfo {author}
  {\bibfnamefont {A.}~\bibnamefont {Brieussel}}, \bibinfo {author}
  {\bibfnamefont {N.}~\bibnamefont {Somaschi}}, \bibinfo {author}
  {\bibfnamefont {H.}~\bibnamefont {Huet}}, \bibinfo {author} {\bibfnamefont
  {A.}~\bibnamefont {Harouri}}, \bibinfo {author} {\bibfnamefont
  {A.}~\bibnamefont {Lema{\^\i}tre}}, \bibinfo {author} {\bibfnamefont
  {I.}~\bibnamefont {Sagnes}}, \bibinfo {author} {\bibfnamefont
  {N.}~\bibnamefont {Belabas}}, \bibinfo {author} {\bibfnamefont
  {F.}~\bibnamefont {Sciarrino}}, \bibinfo {author} {\bibfnamefont
  {R.}~\bibnamefont {Osellame}}, \bibinfo {author} {\bibfnamefont
  {P.}~\bibnamefont {Senellart}},\ and\ \bibinfo {author} {\bibfnamefont
  {A.}~\bibnamefont {Crespi}},\ }\href
  {https://doi.org/10.1103/PhysRevX.12.031033} {\bibfield  {journal} {\bibinfo
  {journal} {Physical Review X}\ }\textbf {\bibinfo {volume} {12}},\ \bibinfo
  {pages} {031033} (\bibinfo {year} {2022})}\BibitemShut {NoStop}%
\bibitem [{\citenamefont {Pont}\ \emph {et~al.}(2024)\citenamefont {Pont},
  \citenamefont {Corrielli}, \citenamefont {Fyrillas}, \citenamefont {Agresti},
  \citenamefont {Carvacho}, \citenamefont {Maring}, \citenamefont {Emeriau},
  \citenamefont {Ceccarelli}, \citenamefont {Albiero}, \citenamefont
  {Ferreira}, \citenamefont {Somaschi}, \citenamefont {Senellart},
  \citenamefont {Sagnes}, \citenamefont {Morassi}, \citenamefont
  {Lema{\^\i}tre}, \citenamefont {Senellart}, \citenamefont {Sciarrino},
  \citenamefont {Liscidini}, \citenamefont {Belabas},\ and\ \citenamefont
  {Osellame}}]{pont2022high}%
  \BibitemOpen
  \bibfield  {author} {\bibinfo {author} {\bibfnamefont {M.}~\bibnamefont
  {Pont}}, \bibinfo {author} {\bibfnamefont {G.}~\bibnamefont {Corrielli}},
  \bibinfo {author} {\bibfnamefont {A.}~\bibnamefont {Fyrillas}}, \bibinfo
  {author} {\bibfnamefont {I.}~\bibnamefont {Agresti}}, \bibinfo {author}
  {\bibfnamefont {G.}~\bibnamefont {Carvacho}}, \bibinfo {author}
  {\bibfnamefont {N.}~\bibnamefont {Maring}}, \bibinfo {author} {\bibfnamefont
  {P.-E.}\ \bibnamefont {Emeriau}}, \bibinfo {author} {\bibfnamefont
  {F.}~\bibnamefont {Ceccarelli}}, \bibinfo {author} {\bibfnamefont
  {R.}~\bibnamefont {Albiero}}, \bibinfo {author} {\bibfnamefont {P.~H.}\
  \bibnamefont {Ferreira}}, \bibinfo {author} {\bibfnamefont {N.}~\bibnamefont
  {Somaschi}}, \bibinfo {author} {\bibfnamefont {J.}~\bibnamefont {Senellart}},
  \bibinfo {author} {\bibfnamefont {I.}~\bibnamefont {Sagnes}}, \bibinfo
  {author} {\bibfnamefont {M.}~\bibnamefont {Morassi}}, \bibinfo {author}
  {\bibfnamefont {A.}~\bibnamefont {Lema{\^\i}tre}}, \bibinfo {author}
  {\bibfnamefont {P.}~\bibnamefont {Senellart}}, \bibinfo {author}
  {\bibfnamefont {F.}~\bibnamefont {Sciarrino}}, \bibinfo {author}
  {\bibfnamefont {M.}~\bibnamefont {Liscidini}}, \bibinfo {author}
  {\bibfnamefont {N.}~\bibnamefont {Belabas}},\ and\ \bibinfo {author}
  {\bibfnamefont {R.}~\bibnamefont {Osellame}},\ }\href
  {https://doi.org/10.1038/s41534-024-00830-z} {\bibfield  {journal} {\bibinfo
  {journal} {npj Quantum Information}\ }\textbf {\bibinfo {volume} {10}},\
  \bibinfo {pages} {50} (\bibinfo {year} {2024})}\BibitemShut {NoStop}%
\bibitem [{\citenamefont {Corrielli}\ \emph {et~al.}(2021)\citenamefont
  {Corrielli}, \citenamefont {Crespi},\ and\ \citenamefont
  {Osellame}}]{corrielli2021femtosecond}%
  \BibitemOpen
  \bibfield  {author} {\bibinfo {author} {\bibfnamefont {G.}~\bibnamefont
  {Corrielli}}, \bibinfo {author} {\bibfnamefont {A.}~\bibnamefont {Crespi}},\
  and\ \bibinfo {author} {\bibfnamefont {R.}~\bibnamefont {Osellame}},\ }\href
  {https://doi.org/10.1515/nanoph-2021-0419} {\bibfield  {journal} {\bibinfo
  {journal} {Nanophotonics}\ }\textbf {\bibinfo {volume} {10}},\ \bibinfo
  {pages} {3789} (\bibinfo {year} {2021})}\BibitemShut {NoStop}%
\bibitem [{\citenamefont {Clements}\ \emph {et~al.}(2016)\citenamefont
  {Clements}, \citenamefont {Humphreys}, \citenamefont {Metcalf}, \citenamefont
  {Kolthammer},\ and\ \citenamefont {Walmsley}}]{clements2016optimal}%
  \BibitemOpen
  \bibfield  {author} {\bibinfo {author} {\bibfnamefont {W.~R.}\ \bibnamefont
  {Clements}}, \bibinfo {author} {\bibfnamefont {P.~C.}\ \bibnamefont
  {Humphreys}}, \bibinfo {author} {\bibfnamefont {B.~J.}\ \bibnamefont
  {Metcalf}}, \bibinfo {author} {\bibfnamefont {W.~S.}\ \bibnamefont
  {Kolthammer}},\ and\ \bibinfo {author} {\bibfnamefont {I.~A.}\ \bibnamefont
  {Walmsley}},\ }\href {https://doi.org/10.1364/OPTICA.3.001460} {\bibfield
  {journal} {\bibinfo  {journal} {Optica}\ }\textbf {\bibinfo {volume} {3}},\
  \bibinfo {pages} {1460} (\bibinfo {year} {2016})}\BibitemShut {NoStop}%
\bibitem [{\citenamefont {Pentangelo}\ \emph {et~al.}(2024)\citenamefont
  {Pentangelo}, \citenamefont {Di~Giano}, \citenamefont {Piacentini},
  \citenamefont {Arpe}, \citenamefont {Ceccarelli}, \citenamefont {Crespi},\
  and\ \citenamefont {Osellame}}]{pentangelo2024high}%
  \BibitemOpen
  \bibfield  {author} {\bibinfo {author} {\bibfnamefont {C.}~\bibnamefont
  {Pentangelo}}, \bibinfo {author} {\bibfnamefont {N.}~\bibnamefont
  {Di~Giano}}, \bibinfo {author} {\bibfnamefont {S.}~\bibnamefont
  {Piacentini}}, \bibinfo {author} {\bibfnamefont {R.}~\bibnamefont {Arpe}},
  \bibinfo {author} {\bibfnamefont {F.}~\bibnamefont {Ceccarelli}}, \bibinfo
  {author} {\bibfnamefont {A.}~\bibnamefont {Crespi}},\ and\ \bibinfo {author}
  {\bibfnamefont {R.}~\bibnamefont {Osellame}},\ }\href
  {https://doi.org/10.1515/nanoph-2023-0636} {\bibfield  {journal} {\bibinfo
  {journal} {Nanophotonics}\ }\textbf {\bibinfo {volume} {13}},\ \bibinfo
  {pages} {2259} (\bibinfo {year} {2024})}\BibitemShut {NoStop}%
\bibitem [{\citenamefont {Flamini}\ \emph {et~al.}(2015)\citenamefont
  {Flamini}, \citenamefont {Magrini}, \citenamefont {Rab}, \citenamefont
  {Spagnolo}, \citenamefont {D'ambrosio}, \citenamefont {Mataloni},
  \citenamefont {Sciarrino}, \citenamefont {Zandrini}, \citenamefont {Crespi},
  \citenamefont {Ramponi},\ and\ \citenamefont
  {Osellame}}]{flamini2015thermally}%
  \BibitemOpen
  \bibfield  {author} {\bibinfo {author} {\bibfnamefont {F.}~\bibnamefont
  {Flamini}}, \bibinfo {author} {\bibfnamefont {L.}~\bibnamefont {Magrini}},
  \bibinfo {author} {\bibfnamefont {A.~S.}\ \bibnamefont {Rab}}, \bibinfo
  {author} {\bibfnamefont {N.}~\bibnamefont {Spagnolo}}, \bibinfo {author}
  {\bibfnamefont {V.}~\bibnamefont {D'ambrosio}}, \bibinfo {author}
  {\bibfnamefont {P.}~\bibnamefont {Mataloni}}, \bibinfo {author}
  {\bibfnamefont {F.}~\bibnamefont {Sciarrino}}, \bibinfo {author}
  {\bibfnamefont {T.}~\bibnamefont {Zandrini}}, \bibinfo {author}
  {\bibfnamefont {A.}~\bibnamefont {Crespi}}, \bibinfo {author} {\bibfnamefont
  {R.}~\bibnamefont {Ramponi}},\ and\ \bibinfo {author} {\bibfnamefont
  {R.}~\bibnamefont {Osellame}},\ }\href {https://doi.org/10.1038/lsa.2015.127}
  {\bibfield  {journal} {\bibinfo  {journal} {Light: Science \& Applications}\
  }\textbf {\bibinfo {volume} {4}},\ \bibinfo {pages} {e354} (\bibinfo {year}
  {2015})}\BibitemShut {NoStop}%
\bibitem [{\citenamefont {Ceccarelli}\ \emph {et~al.}(2019)\citenamefont
  {Ceccarelli}, \citenamefont {Atzeni}, \citenamefont {Prencipe}, \citenamefont
  {Farinaro},\ and\ \citenamefont {Osellame}}]{ceccarelli2019thermal}%
  \BibitemOpen
  \bibfield  {author} {\bibinfo {author} {\bibfnamefont {F.}~\bibnamefont
  {Ceccarelli}}, \bibinfo {author} {\bibfnamefont {S.}~\bibnamefont {Atzeni}},
  \bibinfo {author} {\bibfnamefont {A.}~\bibnamefont {Prencipe}}, \bibinfo
  {author} {\bibfnamefont {R.}~\bibnamefont {Farinaro}},\ and\ \bibinfo
  {author} {\bibfnamefont {R.}~\bibnamefont {Osellame}},\ }\href
  {https://doi.org/10.1109/JLT.2019.2923126} {\bibfield  {journal} {\bibinfo
  {journal} {Journal of Lightwave Technology}\ }\textbf {\bibinfo {volume}
  {37}},\ \bibinfo {pages} {4275} (\bibinfo {year} {2019})}\BibitemShut
  {NoStop}%
\bibitem [{\citenamefont {Ceccarelli}\ \emph {et~al.}(2020)\citenamefont
  {Ceccarelli}, \citenamefont {Atzeni}, \citenamefont {Pentangelo},
  \citenamefont {Pellegatta}, \citenamefont {Crespi},\ and\ \citenamefont
  {Osellame}}]{ceccarelli2020low}%
  \BibitemOpen
  \bibfield  {author} {\bibinfo {author} {\bibfnamefont {F.}~\bibnamefont
  {Ceccarelli}}, \bibinfo {author} {\bibfnamefont {S.}~\bibnamefont {Atzeni}},
  \bibinfo {author} {\bibfnamefont {C.}~\bibnamefont {Pentangelo}}, \bibinfo
  {author} {\bibfnamefont {F.}~\bibnamefont {Pellegatta}}, \bibinfo {author}
  {\bibfnamefont {A.}~\bibnamefont {Crespi}},\ and\ \bibinfo {author}
  {\bibfnamefont {R.}~\bibnamefont {Osellame}},\ }\href
  {https://doi.org/10.1002/lpor.202000024} {\bibfield  {journal} {\bibinfo
  {journal} {Laser \& Photonics Reviews}\ }\textbf {\bibinfo {volume} {14}},\
  \bibinfo {pages} {2000024} (\bibinfo {year} {2020})}\BibitemShut {NoStop}%
\bibitem [{\citenamefont {Albiero}\ \emph {et~al.}(2022)\citenamefont
  {Albiero}, \citenamefont {Pentangelo}, \citenamefont {Gardina}, \citenamefont
  {Atzeni}, \citenamefont {Ceccarelli},\ and\ \citenamefont
  {Osellame}}]{albiero2022}%
  \BibitemOpen
  \bibfield  {author} {\bibinfo {author} {\bibfnamefont {R.}~\bibnamefont
  {Albiero}}, \bibinfo {author} {\bibfnamefont {C.}~\bibnamefont {Pentangelo}},
  \bibinfo {author} {\bibfnamefont {M.}~\bibnamefont {Gardina}}, \bibinfo
  {author} {\bibfnamefont {S.}~\bibnamefont {Atzeni}}, \bibinfo {author}
  {\bibfnamefont {F.}~\bibnamefont {Ceccarelli}},\ and\ \bibinfo {author}
  {\bibfnamefont {R.}~\bibnamefont {Osellame}},\ }\href
  {https://doi.org/10.3390/mi13071145} {\bibfield  {journal} {\bibinfo
  {journal} {Micromachines}\ }\textbf {\bibinfo {volume} {13}},\ \bibinfo
  {pages} {1145} (\bibinfo {year} {2022})}\BibitemShut {NoStop}%
\bibitem [{\citenamefont {Ceccarelli}\ \emph {et~al.}(2024)\citenamefont
  {Ceccarelli}, \citenamefont {Rakonjac}, \citenamefont {Grandi}, \citenamefont
  {de~Riedmatten}, \citenamefont {Osellame},\ and\ \citenamefont
  {Corrielli}}]{ceccarelli2024integrated}%
  \BibitemOpen
  \bibfield  {author} {\bibinfo {author} {\bibfnamefont {F.}~\bibnamefont
  {Ceccarelli}}, \bibinfo {author} {\bibfnamefont {J.~V.}\ \bibnamefont
  {Rakonjac}}, \bibinfo {author} {\bibfnamefont {S.}~\bibnamefont {Grandi}},
  \bibinfo {author} {\bibfnamefont {H.}~\bibnamefont {de~Riedmatten}}, \bibinfo
  {author} {\bibfnamefont {R.}~\bibnamefont {Osellame}},\ and\ \bibinfo
  {author} {\bibfnamefont {G.}~\bibnamefont {Corrielli}},\ }\href
  {https://doi.org/10.1088/2515-7647/ad82c2} {\bibfield  {journal} {\bibinfo
  {journal} {Journal of Physics: Photonics}\ }\textbf {\bibinfo {volume} {6}},\
  \bibinfo {pages} {045023} (\bibinfo {year} {2024})}\BibitemShut {NoStop}%
\bibitem [{\citenamefont {Kok}\ \emph {et~al.}(2007)\citenamefont {Kok},
  \citenamefont {Munro}, \citenamefont {Nemoto}, \citenamefont {Ralph},
  \citenamefont {Dowling},\ and\ \citenamefont {Milburn}}]{kok2007linear}%
  \BibitemOpen
  \bibfield  {author} {\bibinfo {author} {\bibfnamefont {P.}~\bibnamefont
  {Kok}}, \bibinfo {author} {\bibfnamefont {W.~J.}\ \bibnamefont {Munro}},
  \bibinfo {author} {\bibfnamefont {K.}~\bibnamefont {Nemoto}}, \bibinfo
  {author} {\bibfnamefont {T.~C.}\ \bibnamefont {Ralph}}, \bibinfo {author}
  {\bibfnamefont {J.~P.}\ \bibnamefont {Dowling}},\ and\ \bibinfo {author}
  {\bibfnamefont {G.~J.}\ \bibnamefont {Milburn}},\ }\href
  {https://doi.org/10.1103/RevModPhys.79.135} {\bibfield  {journal} {\bibinfo
  {journal} {Reviews of Modern Physics}\ }\textbf {\bibinfo {volume} {79}},\
  \bibinfo {pages} {135} (\bibinfo {year} {2007})}\BibitemShut {NoStop}%
\end{thebibliography}


\end{document}

\title{Supplementary Material: Experimental verification of Threshold Quantum State Tomography on a fully-reconfigurable photonic integrated circuit}

\author{Eugenio Caruccio}
\affiliation{Dipartimento di Fisica, Sapienza Universit\`{a} di Roma, Piazzale Aldo Moro 5, I-00185 Roma, Italy}

\author{Diego Maragnano}
\affiliation{Dipartimento di Fisica, Universit\`{a} di Pavia, via Bassi 6, 27100 Pavia, Italy}

\author{Giovanni Rodari}
\affiliation{Dipartimento di Fisica, Sapienza Universit\`{a} di Roma, Piazzale Aldo Moro 5, I-00185 Roma, Italy}

\author{Davide Picus}
\affiliation{Dipartimento di Fisica, Sapienza Universit\`{a} di Roma, Piazzale Aldo Moro 5, I-00185 Roma, Italy}

\author{Giovanni Garberoglio}
\affiliation{European Centre for Theoretical Studies in Nuclear Physics and Related Areas (ECT*, Fondazione Bruno Kessler); Villa Tambosi, Strada delle Tabarelle 286, I-38123 Villazzano (TN), Italy}

\author{Daniele Binosi}
\affiliation{European Centre for Theoretical Studies in Nuclear Physics and Related Areas (ECT*, Fondazione Bruno Kessler); Villa Tambosi, Strada delle Tabarelle 286, I-38123 Villazzano (TN), Italy}

\author{Riccardo Albiero}
\affiliation{Istituto di Fotonica e Nanotecnologie, Consiglio Nazionale delle Ricerche (IFN-CNR), 
Piazza Leonardo da Vinci, 32, I-20133 Milano, Italy}

\author{Niki Di Giano}
\affiliation{Dipartimento di Fisica, Politecnico di Milano, Piazza Leonardo da Vinci 32, 20133 Milano, Italy}
\affiliation{Istituto di Fotonica e Nanotecnologie, Consiglio Nazionale delle Ricerche (IFN-CNR), 
Piazza Leonardo da Vinci, 32, I-20133 Milano, Italy}

\author{Francesco Ceccarelli}
\affiliation{Istituto di Fotonica e Nanotecnologie, Consiglio Nazionale delle Ricerche (IFN-CNR), 
Piazza Leonardo da Vinci, 32, I-20133 Milano, Italy}

\author{Giacomo Corrielli}
\affiliation{Istituto di Fotonica e Nanotecnologie, Consiglio Nazionale delle Ricerche (IFN-CNR), 
Piazza Leonardo da Vinci, 32, I-20133 Milano, Italy}

\author{Nicol\`o Spagnolo}
\affiliation{Dipartimento di Fisica, Sapienza Universit\`{a} di Roma, Piazzale Aldo Moro 5, I-00185 Roma, Italy}

\author{Roberto Osellame}
\email{roberto.osellame@cnr.it}
\affiliation{Istituto di Fotonica e Nanotecnologie, Consiglio Nazionale delle Ricerche (IFN-CNR), 
Piazza Leonardo da Vinci, 32, I-20133 Milano, Italy}

\author{Maurizio Dapor}
\email{dapor@ectstar.eu}
\affiliation{European Centre for Theoretical Studies in Nuclear Physics and Related Areas (ECT*, Fondazione Bruno Kessler); Villa Tambosi, Strada delle Tabarelle 286, I-38123 Villazzano (TN), Italy}

\author{Marco Liscidini}
\email{marco.liscidini@unipv.it}
\affiliation{Dipartimento di Fisica, Universit\`{a} di Pavia, via Bassi 6, 27100 Pavia, Italy}

\author{Fabio Sciarrino}
\email{fabio.sciarrino@uniroma1.it}
\affiliation{Dipartimento di Fisica, Sapienza Universit\`{a} di Roma, Piazzale Aldo Moro 5, I-00185 Roma, Italy}

%---------------------------------------------
\maketitle
%---------------------------------------------

\section{Interferometer programming for state generation}

In this section we discuss the interferometer programming procedure to generate in post-selection different qubits states in the dual rail logic. Let us consider the scenario where $n$ input photons are injected in a $m$-mode linear interferometer, described by a unitary matrix $U$ that defines the input-output relations for the bosonic operators as $b_{j} = \sum_{i} U_{ji} a_{i}$. Let us call $\vert S \rangle = \vert s_{1}, s_{2}, \ldots, s_{m} \rangle$ the generic input state in the Fock representation, where $s_{i}$ is the number of photons in input port $i$.  According to the bosonic transformation rule in a linear optical circuit, the transition amplitude from an input configuration $\vert S \rangle$ to an output configuration $\vert T \rangle$ reads $\beta_{S \rightarrow T} = \mathrm{per}(U_{S,T})/\sqrt{\prod_i s_i! \prod_j t_j!}$, where $U_{S,T}$ is the $n \times n$ matrix obtained by selecting $t_{j}$ times rows $j$ of $U$, and $s_i$ times columns $i$ of $U$. The output state when injecting the interferometer with input state $\vert S \rangle$ will thus be in a superposition of all output configuration as:
\begin{equation}
    \label{eq:linear_optics}
    \vert \psi \rangle = \sum_{T} \beta_{S \rightarrow T} \vert T \rangle.
\end{equation}

When using this bosonic platform to encode $n$ qubits in the dual rail logic, with a $m=2n$-mode interferometer, one needs to post-select to those configurations satisfying the correct condition for the dual rail logic, that is, one photon in each of mode pairs (1,2), (3,4), $\ldots$, ($2n-1,2n$). In our case, the input photons are injected in the odd ports of the interferometer, and thus the input state corresponds to setting $s_{2l-1} = 1$ and $s_{2l} = 0$, with $l=1, \ldots, n$. By calling $\mathrm{DR}$ the set of possible output configurations $T$ satisfying the condition for the dual rail logic, the post-selected output state reads:
\begin{equation}
    \label{eq:post-selected_linear_optics}
    \vert \psi^{\prime} \rangle = \sum_{T \in \mathrm{DR}} c_{T} \vert T \rangle = \sum_{T \in \mathrm{DR}} \frac{\beta_{S \rightarrow T}}{\sqrt{p}} \vert T \rangle = \sum_{T \in \mathrm{DR}} \frac{\mathrm{per}(U_{S,T})}{\sqrt{\prod_i s_i! \prod_j t_j!} \sqrt{\sum_{T \in \mathrm{DR}} \vert \beta_{S \rightarrow T} \vert^{2}}} \vert T \rangle.
\end{equation}
Here, $p$ is the post-selection probability that acts as a normalization condition:
\begin{equation}
p = \sum_{T \in \mathrm{DR}} \vert \beta_{S \rightarrow T} \vert^{2}.
\end{equation}
Thus, it is possible to generate different $n$-qubit states in the dual rail logic according to this procedure. We observe that this scheme relies on multiphoton interference, and thus correct generation of a quantum state is sensitive to the presence of partial photon distinguishability. In our experiment, such a procedure has been used to generate all resource states. 

This approach requires implementation of unitary transformations between the modes, that in our case correspond to the one performed in the first set of six layers of Reconfigurable Beam splitters (RBSs). More specifically, the state preparation layers are used to implement the transformation $U$ by tuning the parameters $\theta, \phi$ of the variable RBSs. The latter are implemented in the device through a Mach-Zehnder interferometer (see also Fig. 2 in the main text), corresponding to the sequence of a phase shift $\phi$ (top mode), a balanced directional coupler, a second phase shift $\theta$ (top mode), and a second balanced directional coupler. This elementary cell is described by a matrix of the form:
\begin{equation}
\label{eq:URBS}
U_{\mathrm{RBS}} = 
\begin{pmatrix} 
e^{\imath \phi} \sin (\theta/2) & \cos (\theta/2) \\ 
e^{\imath \phi} \cos (\theta/2)& - \sin (\theta/2)
\end{pmatrix}.
\end{equation}
Application of a sequence of RBSs according to a specific layout, leads to a $2n \times 2n$ matrix $U = U(\{\theta_{i}\}, \{\phi_{j}\})$. Its use in Eq. (\ref{eq:post-selected_linear_optics}) provides the values of the state coefficients generated in post-selection on the output modes. Analogously, a single layer of RBSs acting on the qubit mode pairs is used to measure the state in the different bases.

\begin{figure*}[ht!]
    \centering
    \includegraphics[width=0.99\textwidth]{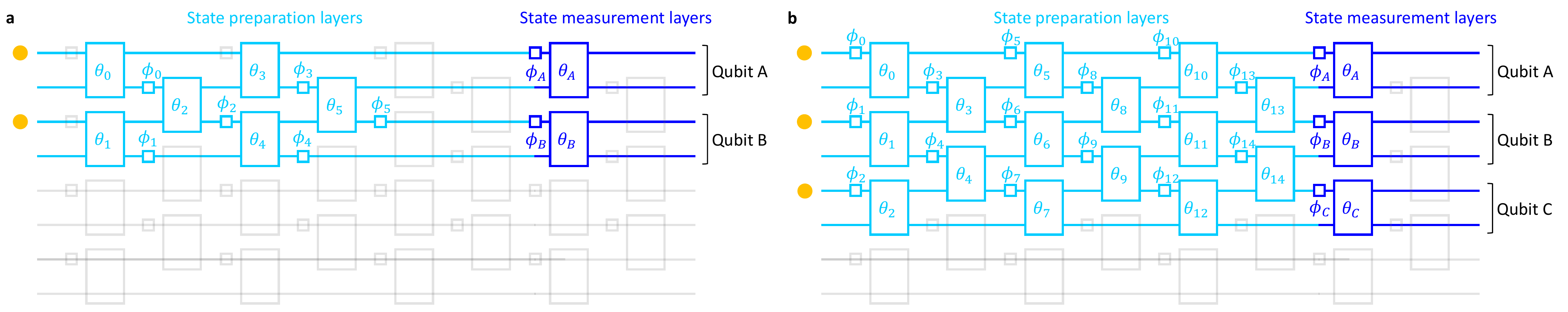}
    \caption{\textbf{Layout of the interferometer programming for the generation of random states with variable Gini index.} Internal scheme to program the device for the generation and measurement of random state of \textbf{a} $n=2$ qubits and \textbf{b} $n=3$. The cyan boxes represent the RBSs which are actually programmed to generate the states, while blue boxes correspond to the elements for the state measurement. Grey boxes are RBSs configured to act as identity, and the grey horizontal lines are unused modes.}
    \label{fig:schemes_Random}
\end{figure*}

We can now discuss the generation of random states with equally-spaced values of the Gini index. The schemes corresponding to how the circuit is programmed are shown in Supplementary Fig. \ref{fig:schemes_Random}. We have then generated a set of random states for both $n=2$ and $n=3$ qubits, by randomly selecting the parameters $\{\theta_{i}\}$ and $\{\phi_{j}\}$, and calculating the corresponding state according to Eq. (\ref{eq:post-selected_linear_optics}). We have then selected 40 (10) states for $n=2$ ($n=3$) qubits with corresponding associated Gini indexes which span with equal spacing the entire possible range for such parameter. This allows to select a set of states with different level of sparsities, and thus test the protocol in different conditions.

\begin{figure*}[ht!]
    \centering
    \includegraphics[width=0.99\textwidth]{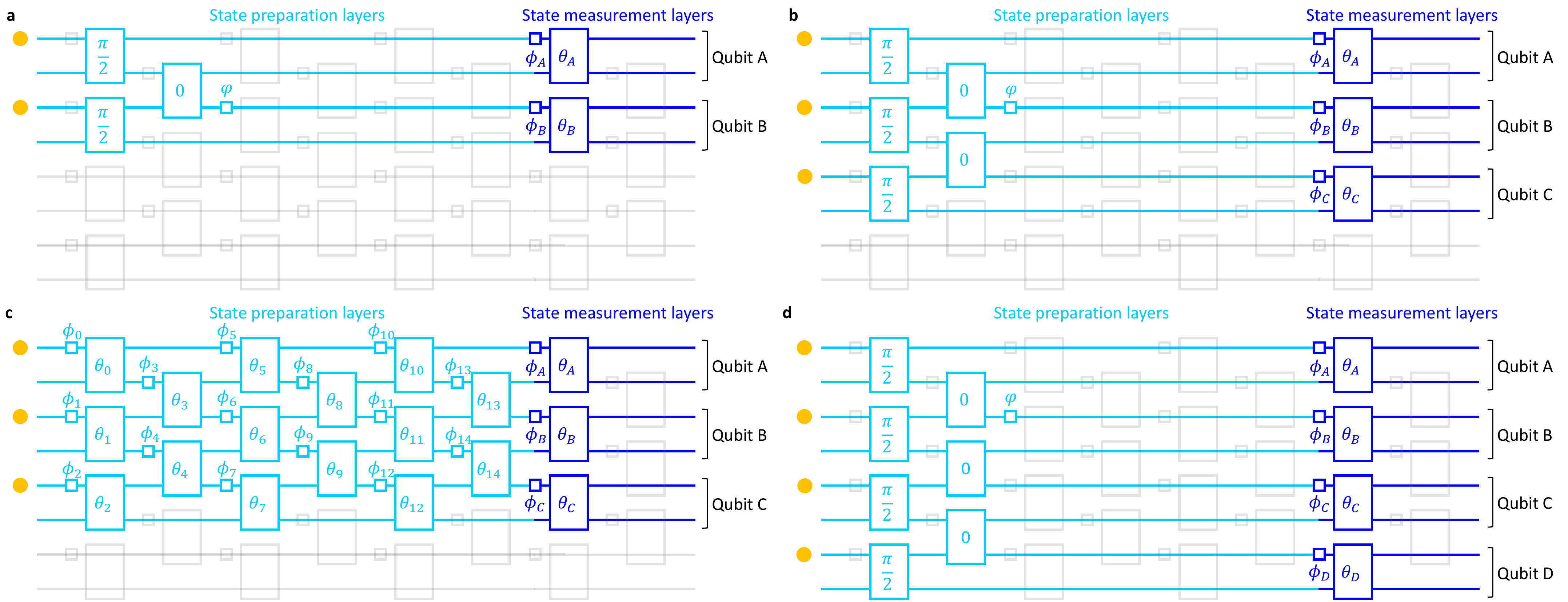}
    \caption{\textbf{Layout of the interferometer programming for the generation of maximally-entangled states.} Internal scheme to program the device for the generation and measurement of maximally-entangled states. \textbf{a}, $\vert \Psi^{+} \rangle$. \textbf{b}, $\vert \mathrm{GHZ}_{3} \rangle$. \textbf{c}, $\vert \mathrm{W}_{3} \rangle$. \textbf{d}, $\vert \mathrm{GHZ}_{4} \rangle$. The cyan boxes represent the RBSs which are actually programmed to generate the states, while blue boxes correspond to the elements for the state measurement. Grey boxes are RBSs configured to act as the identity, and the grey horizontal lines are unused modes. For $\vert \Psi^{+} \rangle$ and for the GHZ states, the values for the RBSs angles $\theta$ are reported in the cyan boxes.}
    \label{fig:schemes_Symm}
\end{figure*}

The same encoding strategy has been used to generate maximally-entangled states, using the layouts reported in Supplementary Fig. \ref{fig:schemes_Symm}. For the Bell state $\vert \Psi^{+} \rangle$ and the GHZ states $\vert \mathrm{GHZ}_{3} \rangle$, $\vert \mathrm{GHZ}_{4} \rangle$, we identify the correct setting for the RBSs in the state preparation layer by following the approach of Ref. \cite{pont2022high}. Generation of this class of states can be performed by setting all RBSs in the first layer to act as balanced beam splitters, while the RBSs in the second layers are all set to act as swap operations between the modes. This configuration, post-selecting to those combinations satisfying the dual rail logic, leads to the generation of the target states. The relative phase between the two superposition terms can be set by tuning a phase $\varphi$ acting in one of the modes. Regarding the $\vert \mathrm{W}_{3} \rangle$ states, its generation requires the full set of 6 generation layers, and an additional set of phase shifts which are inserted by modifying phases $\phi_{A}$ and $\phi_{C}$ at the input of the measurement layers (see Supplementary Tab. \ref{tab:settings_W}).

\begin{table}[ht!]
\begin{tabular}{|c|c|c|c|c|c|c|c|c|c|c|c|c|c|c|}
\hline
\multicolumn{15}{|c|}{Parameters settings for the generation of a $\vert \mathrm{W}_{3} \rangle$ state} \\
\hline
$\theta_{0}$ & $\theta_{1}$ & $\theta_{2}$ & $\theta_{3}$ & $\theta_{4}$ & $\theta_{5}$ & $\theta_{6}$ & $\theta_{7}$ & $\theta_{8}$ & $\theta_{9}$ & $\theta_{10}$ & $\theta_{11}$ & $\theta_{12}$ & $\theta_{13}$ & $\theta_{14}$ \\
\hline
3.141592 & 3.141592 & 1.570796 & 3.141592 & 1.230959 & 1.570796 & 3.141592 & 1.570796 & 1.230959 & 3.141592 & 1.570796 & 0 & 3.141592 & 0 & 0\\
\hline
$\phi_{0}$ & $\phi_{1}$ & $\phi_{2}$ & $\phi_{3}$ & $\phi_{4}$ & $\phi_{5}$ & $\phi_{6}$ & $\phi_{7}$ & $\phi_{8}$ & $\phi_{9}$ & $\phi_{10}$ & $\phi_{11}$ & $\phi_{12}$ & $\phi_{13}$ & $\phi_{14}$ \\
\hline
0 &  0 & 0 & 0 & 3.141592 & 2.094402 & 0 & 4.712389 & 2.094388 & 2.094402 & 0.523596 & 0 & 0& 0 & 0 \\
\hline
\end{tabular}
\caption{\textbf{Settings for the generation of a $\vert \mathrm{W}_{3} \rangle$ state}. Parameters $\theta_{i}$ and $\phi_{i}$ to be inserted in the circuit according to the layout of Supplementary Fig. \ref{fig:schemes_Symm}b. In addition, the state requires two additional phase shifts at modes 1 and 5, before the measurement state. In our experiment, this is performed by shifting the measurement phases $\phi_{A}$ and $\phi_{C}$ of the quantities $\Delta \phi_{A} = 2.618002$, $\Delta \phi_{C} = -2.094406$.}
\label{tab:settings_W}
\end{table}

\section{Modeling experimental imperfections}
\label{sec:model}

In the main text, the retrieved density matrices have been compared with an expectation calculated through a theoretical model that takes into account the main experimental imperfections.

A first class of imperfections in the apparatus arises from two aspects related to the quantum-dot. On a first note, it is necessary to take into account multiphoton emission from the source. This corresponds to having, with a small probability, the presence of two photons in the same temporal bin. In our case, the extra photon has to be attributed to unfiltered residual light from the excitation laser, and is thus fully distinguishable with respects to those emitted from the quantum-dot. This noise process is modeled by considering that, in each time bin, the state is described by an effective density matrix of the form: $\rho = p_0 \vert 0 \rangle \langle 0 \vert + p_1 \vert 1 \rangle \langle 1 \vert + p_2 \vert 1,\tilde{1} \rangle \langle 1, \tilde{1} \vert$, where $\langle 1, \tilde{1} \vert$ stands for the presence of two photons in distinguishable internal states, while $(p_0, p_1, p_2)$ are the probabilities of having respectively 0, 1 or 2 photons in the time bin. In our case, $p_0$ at the emission stage is small, while $p_1$ and $p_2$ can be evaluated from the amount of measured single-photon signal and from the second-order correlation function $g^{(2)}(0) = 2p_2/(p_1+2p_2)^2$. The $g^{(2)}(0)$ value can be retrieved via a Hanbury Brown-Twiss experiment \cite{hbt}. The second imperfection arising from the source is related to partial photon indistinguishability between the emitted photons. This is modeled via the Gram-matrix formalism \cite{tichy2015sampling}. In our case, the photons emitted from the quantum-dot are characterized \cite{rodari2024semi} by a real-valued Gram-matrix, with elements $S_{ij} = \sqrt{M_{ij}}$, where $M_{ij} = \vert \langle \psi_i \vert \psi_j \rangle \vert^2$ are the overlaps between photons $(i,j)$. The overlaps can be estimated from the visibilities $V^{\mathrm{HOM}}_{ij}$ obtained from pairwise Hong-Ou-Mandel experiments \cite{hom1987} between each photon pairs. In the presence of multiphoton emission with additional distinguishable noise photons, the overlaps $M_{ij}$ are obtained from the visibilities, in the limit of low $g^{(2)}(0)$, as $M_{ij}=[V^{\mathrm{HOM}}_{ij} + g^{(2)}(0)]/[1 - g^{(2)}(0)]$ \cite{Olli21}.

The second class of noise effects in the density matrix arises from the integrated photonic processor. In particular, imperfections in the device may lead to the implementation of a different unitary with respect to the one necessary for the generation of the chosen state. As previously discussed, the RBS elementary cell is implemented via a Mach-Zehnder interferometer comprising two balanced directional couplers. Small imperfections in the fabrication process lead to couplers having reflectivities slightly different than 0.5, which in our case are found in the interval [0.50, 0.58]. This affects the minimum and maximum values that the overall reflectivity of the RBS can reach via its programmability. In the ideal scenario, the overall reflectivity $R(\theta)$ of the RBS is found to be [see Eq. (\ref{eq:URBS})] $R(\theta) =  \sin^2(\theta/2)$. Conversely, if the directional couplers of the elementary cell are characterized by reflectivities $r_1$ and $r_2$, the reflectivity of the RBS reads $R(\theta) = r_1 r_2 + t_1 t_2 - 2 \sqrt{r_1 r_2 t_1 t_2} \cos \theta$. Hence, the corresponding minimum values $R_{\mathrm{min}}$ and $R_{\mathrm{max}}$ achievable by tuning $\theta$ are respectively:
\begin{eqnarray}
R_{\mathrm{min}} &=& r_1 r_2 + t_1 t_2 - 2 \sqrt{r_1 r_2 t_1 t_2}, \\
R_{\mathrm{max}} &=& r_1 r_2 + t_1 t_2 + 2 \sqrt{r_1 r_2 t_1 t_2},
\end{eqnarray}
where $t_i = 1 - r_i$. If $r_1, r_2 \neq 0.5$, this prevents reaching values close to $R\sim 0$ and $R\sim 1$ in the RBS, thus adding small errors in the state generation process.

Finally, we have also taken into account losses in the apparatus. In our case, losses are found to be almost balanced between the modes, and can thus be included in the model as a single round of loss after the quantum-dot source \cite{rodari2024semi,Oszm18}. The only relevant contribution of unbalanced losses can be found in the different detection efficiencies of the employed detector, which have been directly corrected in the experimental data via an appropriate renormalization of the measured counts. Indeed, this is not a noise effect in the state generation process, but arises only at the detection/verification stage. 

Having included all these noise effects, the expectations on the density matrices are obtained by the following procedure. For each state to be tested, we calculated the different output probability distributions corresponding to a circuit programming that implements (i) the state generation unitary and (ii) all possible combinations of unitaries corresponding to the measurement of the Pauli operators at the measurement layers. Combining this set of probability distributions leads to a prediction for the state density matrix.

\section{Additional data analyses and figures of merit}

In this Section we report additional analyses for the experimental data reported in the main text. 

First, in Supplementary Tabs. \ref{tab:giniindexes2}-\ref{tab:giniindexes3} we report the Gini indexes associated to these two sets of quantum states. 
\begin{table}[ht!]
\begin{tabular}{|c|c|c|c|c|c|c|c|}
\hline
\multicolumn{8}{|c|}{2-qubit quantum states} \\
\hline
state number & Gini index & state number & Gini index & state number & Gini index & state number & Gini index \\
\hline
1 & 0.000944 & 11 & 0.06313556 & 21 & 0.12627012 & 31 & 0.18807192 \\
2 & 0.00663219 & 12 & 0.06867141 & 22 & 0.13204937 & 32 & 0.19487812 \\
3 & 0.01249306 & 13 & 0.07593002 & 23 & 0.13790891 & 33 & 0.20180418\\
4 & 0.01814765 & 14 & 0.0811412 & 24 & 0.14387539 & 34 & 0.20715015 \\
5 & 0.02421449 & 15 & 0.08702726 & 25 & 0.15024978 & 35 & 0.21351848\\
6 & 0.03227125 & 16 & 0.09500147 & 26 & 0.1567441 & 36 & 0.22003493\\
7 & 0.03700212 & 17 & 0.09970305 & 27 & 0.1639128 & 37 & 0.22618367\\
8 & 0.04446993 & 18 & 0.10596755 & 28 & 0.16907807 & 38 &  0.23185384\\
9 & 0.05089243 & 19 & 0.11257372 & 29 & 0.17615138 & 39 & 0.23896472\\
10 & 0.05707232& 20 & 0.11851598 & 30 & 0.1824406 & 40 & 0.24496395\\
\hline
\end{tabular}
\caption{\textbf{Gini indexes for the random 2-qubit states}. Value of the Gini index for the different 2-qubit states chosen in the reported experiment. The reported Gini index corresponds to the ideal state, thus in the absence of experimental imperfections.}
\label{tab:giniindexes2}
\end{table}
\begin{table}[ht!]
\begin{tabular}{|c|c|c|c|c|c|c|c|}
\hline
\multicolumn{2}{|c|}{3-qubit quantum states} \\
\hline
state number & Gini index \\
\hline
1 & 0.0078638  \\
2 & 0.02123207 \\
3 & 0.03362007 \\
4 & 0.04380996 \\
5 & 0.05467637 \\
6 & 0.06720037 \\
7 & 0.08189606 \\
8 & 0.09318023 \\
9 & 0.10441946 \\
10 & 0.11615141 \\
\hline
\end{tabular}
\caption{\textbf{Gini indexes for the random 3-qubit states}. Value of the Gini index for the different 3-qubit states chosen in the reported experiment. The reported Gini index corresponds to the ideal state, thus in the absence of experimental imperfections}
\label{tab:giniindexes3}
\end{table}
In Supplementary Fig. \ref{fig:fidelitiestqst_random} we then perform an additional analysis on the $2-$ and $3-$ qubits random states. 
\begin{figure*}[ht!]
    \centering
    \includegraphics[width=0.99\textwidth]{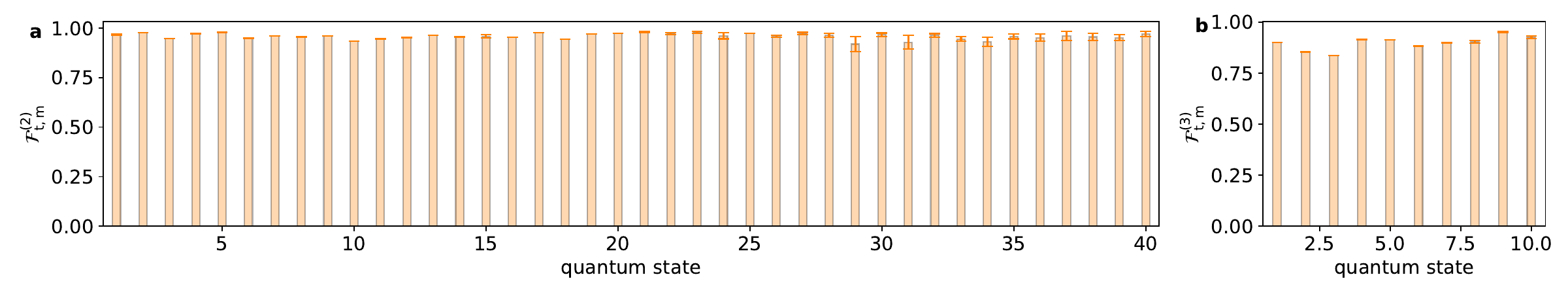}
    \caption{\textbf{Additional results of the state reconstruction process for random states.} \textbf{a}, Scenario with $n=2$ tested with 40 different states. \textbf{b}, Scenario with $n=3$ tested with 10 different states. In the plots, we report the fidelities $\mathcal{F}^{(n)}_{\mathrm{t,m}}$ (with $n=2,3$) between the state reconstructed via tQST choosing the threshold according to the Gini index and the theoretical model taking into account experimental imperfections (orange bars). The states are ordered on the $x$-axis according to the associated Gini indexes, which are equally-spaced in the range between its minimum and maximum values.}
    \label{fig:fidelitiestqst_random}
\end{figure*}
More specifically, we show the fidelities $\mathcal{F}^{(n)}_{\mathrm{t,m}}$ between the reconstructed states via tQST, using the threshold chosen according to the Gini index, and the expected state calculated with the model described above. 
\begin{figure*}[ht!]
    \centering
    \includegraphics[width=0.99\textwidth]{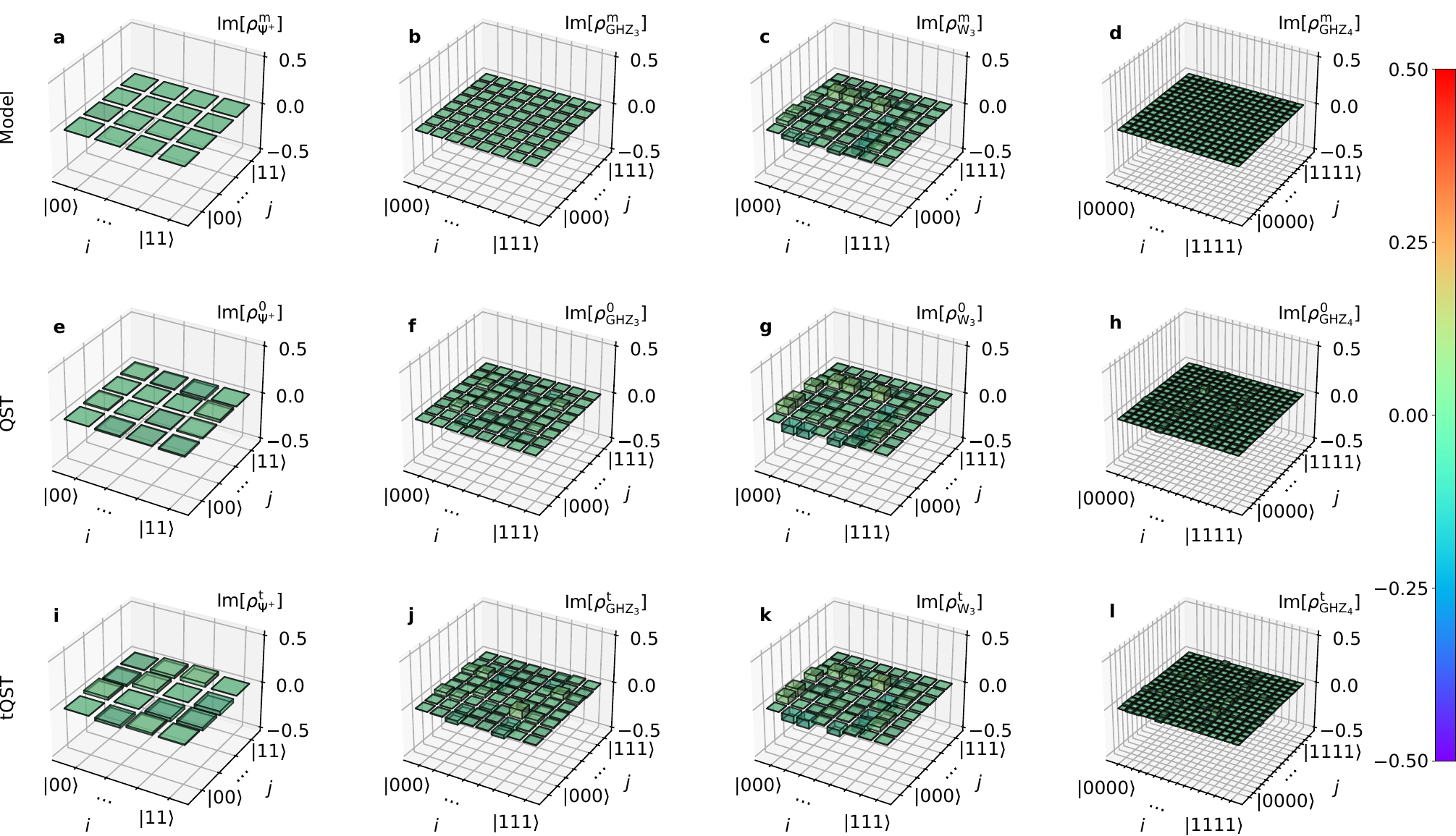}
    \caption{\textbf{Additional results of the state reconstruction process for maximally-entangled states.} Imaginary part of the density matrices for the different tested maximally-entangled states, namely $\vert \Psi^{+} \rangle$ for $n=2$, $\vert \mathrm{GHZ}_{3} \rangle$ and $\vert \mathrm{W}_{3} \rangle$ for $n=3$ and $\vert \mathrm{GHZ}_{4} \rangle$ for $n=4$ (from left to right). In the first row (panels \textbf{a-d}) we report the expected density matrices ($\rho^{\mathrm{m}}_{\alpha}$) estimated from the model taking into account experimental imperfections. The second row (panels \textbf{e-h}) reports the corresponding experimentally reconstructed density matrices ($\rho^{0}_{\alpha}$) via the QST approach, while the third row (panels \textbf{i-l}) reports the density matrices ($\rho^{\mathrm{t}}_{\alpha}$) retrieved via the tQST approach with threshold chosen according to the Gini index. The index $\alpha$ labels the states as $\alpha = \Psi^{+}, \mathrm{GHZ}_{3}, \mathrm{W}_{3}, \mathrm{GHZ}_{4}$. On the right part of the figure, we report the colormap for the density matrix bars, equal for all plots \textbf{a}-\textbf{l}.}
    \label{fig:densitymatrices_SI}
\end{figure*}
The average values of these fidelities over the tested states are respectively $\langle \mathcal{F}^{(2)}_{\mathrm{t,m}} \rangle =0.959\pm0.018$ for the $2$-qubit case, and $\langle \mathcal{F}^{(3)}_{\mathrm{t,m}} \rangle = 0.899\pm 0.033$, for the $3$-qubit scenario. Since these values are compatible with those obtained via QST, such analysis adds an additional supporting element in favor of the observation that the tQST approach does not introduce  significant loss of information on the measured states.

Then, we discuss further analysis on the application of the tQST approach to maximally-entangled states. In Supplementary Fig. \ref{fig:densitymatrices_SI} we show, for the sake of completeness, the imaginary parts of the reconstructed density matrices (the real parts are shown in the main text). Given that all states have, in the ideal scenario, real valued density matrices, the obtained imaginary parts are small. The slight deviations with respect to having real-valued density matrices must be attributed to experimental imperfections in the unitary transformations for both state generation and measurements stages. Finally, as for the random states, we complement the analysis reported in the main text by showing in Supplementary Fig. \ref{fig:fidelitiespurities_SI} the fidelities $\mathcal{F}^{(\alpha)}_{\mathrm{t,m}}$ between the states reconstructed via tQST and the states expected from the model, as a function of the number of projectors $\mathcal{N}$. 

\begin{figure*}[ht!]
    \centering
    \includegraphics[width=0.99\textwidth]{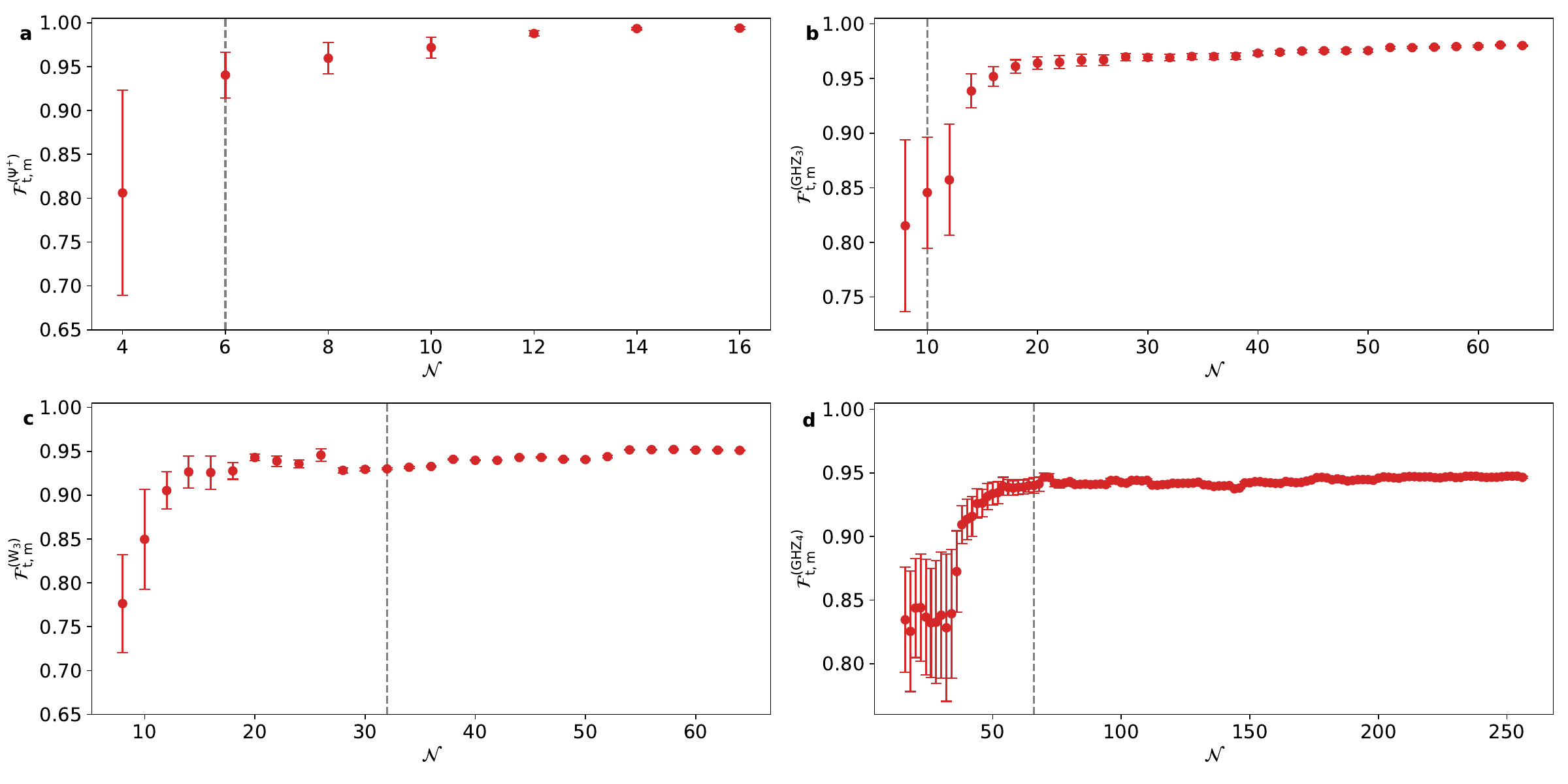}
    \caption{\textbf{Additional figures of merits on the reconstruction of maximally-entangled states.} Plots of the fidelities ($\mathcal{F}^{(\alpha)}_{\mathrm{t,m}}$) between the state reconstructed via tQST, and the expected state for the model. The index $\alpha$ labels the states as $\alpha = \Psi^{+}, \mathrm{GHZ}_{3}, \mathrm{W}_{3}, \mathrm{GHZ}_{4}$. \textbf{a}, Plots for state $\vert \Psi^{+} \rangle$. \textbf{b}, Plots for state $\vert \mathrm{GHZ}_{3} \rangle$. \textbf{c}, Plots for state $\vert \mathrm{W}_{3} \rangle$. \textbf{d}, Plots for state $\vert \mathrm{GHZ}_{4} \rangle$. In all plots, the vertical dashed lines correspond to the number of projectors obtained using the threshold computed from the Gini index, as in Eq.~(2) of the main text.}
    \label{fig:fidelitiespurities_SI}
\end{figure*}

%\bibliography{main.bib}